\documentstyle[seceq,supplement,epsf]{ptptex}

\newcommand{\simgt}{\lower.5ex\hbox{$\; \buildrel > \over \sim \;$}}
\newcommand{\simlt}{\lower.5ex\hbox{$\; \buildrel < \over \sim \;$}}
\newcommand{\himpc}{{\hbox {$h^{-1}$}{\rm Mpc}} }
\newcommand{\sperp}{{\scriptscriptstyle\perp}}
\newcommand{\spara}{{\scriptscriptstyle\parallel}}
\newcommand{\A}{{\scriptscriptstyle A}}
\newcommand{\mass}{{\rm\scriptscriptstyle mass}}
\newcommand{\zm}{z_{\rm max}}
\newcommand{\bfx}{{\bf x}}
\newcommand{\bft}{{\vec\gamma}}
\newcommand{\nLC}{{n^{\rm LC}}}

\newcommand{\rmax}{{r_{\rm max}}}
\newcommand{\rmin}{{r_{\rm min}}}

\newcommand{\bfs}{{\mbox{\boldmath $s$}}}
\newcommand{\bfR}{{\bf R}}
\newcommand{\bfk}{{\bf k}}

\newcommand{\NL}{ {\scriptscriptstyle {NL}} }
\renewcommand{\L}{ {\scriptscriptstyle {L}} }
\pubinfo{Vol. 133, 1999}  %Editorial Office use
\notypesetlogo  %comment in if to eliminate PTPTeX logo
\title{%        %You can use \\ for explicit line-break
Deciphering Cosmological Information 
\\
from Redshift Surveys of High-$z$ Objects
}
\subtitle{ The Cosmological Light-Cone Effect and Redshift-Space Distortion
}    %use this when you want a subtitle

\author{%       %Use \sc for the family name
Yasushi {\sc Suto}, Hiromitsu {\sc Magira}, Y.~P. {\sc Jing},
\\ Takahiko {\sc Matsubara}$^{*}$ and Kazuhiro {\sc Yamamoto}$^{**}$
}

\inst{%         %Affiliation, neglected when [addenda] or [errata]
Department of Physics and Research Center for
    the Early Universe \\ School of Science, University of
    Tokyo, Tokyo 113-0033, Japan
\\
$^{*}$ Department of Physics and Astronomy, 
The Johns Hopkins University
\\ 3400 North Charles Street,
Baltimore, MD 21218-2686, USA
\\
$^{**}$ Department of Physics, Hiroshima
    University \\ Higashi-Hiroshima 739-8526, Japan
}

\recdate{February 17, 1999}

\abst{ The three-dimensional distribution of astronomical objects
observed in redshift space significantly differs from the true
distribution, since the distance to each object cannot be determined by
its redshift $z$ only; for $z \ll 1$ the peculiar velocity field
contaminates the true recession velocity of the Hubble flow, while the
true distance for objects with $z\simgt 1$ sensitively depends on the
(unknown and thus assumed) cosmological parameters. This hampers the
effort to understand the true distribution of the large-scale structure of
the universe. In addition, all cosmological observations are carried out
on a light-cone, the null hypersurface of an observer at $z=0$.  This
implies that their intrinsic properties and clustering statistics should
change even within the survey volume. Therefore, a proper comparison
taking account of the light-cone effect is important to extract any
cosmological information from redshift catalogues, especially for
$z\simgt 1$. We present recent theoretical developments on the two effects
-- the cosmological light-cone effect and the cosmological
redshift-space distortion -- that should play key roles in
observational cosmology in the 21st century.  }

\begin{document}

\maketitle

\section{Introduction}

Galaxy redshift surveys in 1980s revealed and established the
existence of large-scale structure\cite{GH} extending around $\sim
100$Mpc in the current universe at $z=0$.  Theoretically, many
cosmological models are known to be more or less successful in
reproducing the structure at redshift $z\sim0$. In fact, however, this
may be largely because there are still several degrees of freedom or
{\it cosmological parameters} needed to appropriately {\it fit} the
observations at $z\sim0$, including the density parameter, $\Omega_0$,
the mass fluctuation amplitude at the top-hat window radius of
$8\himpc$, $\sigma_8$, the Hubble constant in units of
$100$km/sec/Mpc, $h$, and even the cosmological constant $\lambda_0$.
This kind of {\it degeneracy} in cosmological parameters among viable
models can be broken by combining the data at higher $z$.

With the on-going redshift surveys of millions of galaxies and quasars
and with large telescopes with high spectral resolution, one can probe
directly the epoch of galaxy formation. One of the most important
goals of cosmology in the next century is to construct a physical
model of galaxy formation and evolution in the observationally
determined cosmological context.  To this time this process has been simply
parameterized by the notorious bias parameter $b$, whatever its
meaning might be.
Presently, many theoretical and observational attempts are in progress to
replace the parameter $b$ by another physical model. Naturally,
observational explorations of the larger-scale structure at $z=0$ and
higher redshifts provide important clues to understanding the origin
of structure in the universe.

Redshift surveys of galaxies definitely serve as the central database
for observational cosmology. In addition to the existing catalogues
including CfA1, CfA2, SSRS, and the Las Campanas survey, upcoming
surveys such as 2dF and SDSS are expected to provide important clues
to our universe. In addition to those {\it shallower} surveys,
clustering in the universe in the range $z= 1 - 3$ has been partially
revealed by, for instance, the Lyman-break galaxies\cite{Steidel} and
X-ray selected AGNs.\cite{Carrera} \ In particular, the
2dF\cite{Boyle} (2-degree Field Survey) and SDSS (Sloan Digital Sky
Survey) QSO redshift surveys promise to extend the observable scale of
the universe by an order of magnitude, up to a few Gpc. A proper
interpretation of such redshift surveys in terms of the clustering
evolution, however, requires an understanding of many cosmological
effects which can be neglected and thus have not been considered
seriously in redshift surveys of $z\ll 1$ objects.

This paper consists of two topics which should play key roles in
the theoretical interpretation of the future redshift surveys of
high-redshift objects, the cosmological light-cone effect (\S
\ref{sec:lightcone}) and redshift-space distortion (\S
\ref{sec:cosred}). Primarily, we intend to review and describe the two
effects in a systematic and comprehensive manner on the basis of 
several of our papers.\cite{YS}\tocite{NMS} \ In addition,
however, we input new materials in \S \ref{subsec:alphacrd} and \S
\ref{subsec:nbodycrd}. Also, \S \ref{subsec:predictxi} presents
theoretical predictions based on a different bias model from that
adopted in one of our previous studies.\cite{YS}

In this spirit, the remainder of this paper is organized as follows.
Section \ref{subsec:defxi} briefly outlines a theoretical formulation
of the two-point correlation function on the light-cone hypersurface,
following Ref.~\citen{YS}. The corresponding theoretical predictions
are presented in Section \ref{subsec:predictxi} with future QSO
redshift surveys in mind. The predictions are based on a different
model for evolution of bias from that adopted in Ref.~\citen{YS}. Thus
they illustrate the extent to which the effect of bias changes the
observable clustering of high-redshift objects. Section
\ref{subsec:highlc} summarizes the light-cone effect on the
higher-order clustering statistics following Ref.~\citen{MSS}.
Section \ref{sec:cosred} starts with the basic idea of the
cosmological redshift-space distortion (\S \ref{subsec:ideacrd}) and
its formulation in linear theory, both of which are on the basis of
Ref.~\citen{MS96}. Then we comment on the systematic bias in
estimating the cosmological parameter from shallower ($z \simlt 0.2$)
galaxy redshift surveys, following  Ref.~\citen{NMS}. The next two
subsections are entirely new; Section \ref{subsec:alphacrd} considers
uncertainties due to the distance formulae in inhomogeneous
cosmological models and to the evolution model of bias, and Section
\ref{subsec:nbodycrd} examines the feasibility of the cosmological
redshift-space distortion as a cosmological test to probe $\Omega_0$
and $\lambda_0$. In the latter we fully explore the nonlinear effects
also using high-resolution $N$-body simulations. Finally, we summarize
the main conclusions in Section \ref{sec:final}.

\section{Cosmological light-cone effect \label{sec:lightcone}}

Observing a distant patch of the universe is equivalent to observing
the past.  Due to the finite light velocity, a line-of-sight direction
of a redshift survey is along the time, as well as spatial, coordinate
axis. Therefore the entire sample does not consist of objects on a
constant-time hypersurface, but rather on a light-cone, i.e., a null
hypersurface defined by observers at $z=0$. This implies that many
properties of the objects change across the depth of the survey volume,
including the mean density, the amplitude of spatial clustering of
dark matter, the bias of luminous objects with respect to mass,
and the intrinsic evolution of the absolute magnitude and spectral
energy distribution. These aspects should be properly taken into
account in order to extract cosmological information from observed
samples of redshift surveys.

For the CfA galaxy survey,\cite{GH} for instance, the survey depth
extends up to a recession velocity of 15000 km/s, which is interpreted
as either $d_{\rm max} = 150\himpc$ in spatial distance or $z_{\rm
  max} = 0.05$ in time difference. This translates to a $\sim 10$\%
difference in the amplitude of $\xi$ and $P(k)$ in linear theory.
Compared with the statistical error of the available sample, this
level of systematic effect is negligible. Thus it is quite common to
compare the observed $\xi$ with the theoretical predictions at $z=0$.
The situation will be entirely different for the upcoming galaxy and
QSO redshift surveys, 2dF and SDSS; $0.3 - 1$ million galaxies up to
$z_{\rm max} = 0.2$, and $0.3 - 1 \times 10^5$ QSOs up to $z_{\rm max}
= 3 - 5$. Such observational samples motivate us to formulate a theory
to describe the clustering statistics, fully incorporating the
light-cone effect. In the remainder of this section, we present
theoretical predictions for two-point\cite{YS} and higher-order
correlation\cite{MSS} functions which are properly defined on the
light-cone.

\subsection{Defining two-point correlation functions on a light-cone 
\label{subsec:defxi}}

In this subsection, we derive an expression for the two-point
correlation function on the light-cone hypersurface in the
spatially-flat Friedmann -- Robertson -- Walker space-time for
simplicity; the line element is given in terms of the conformal time
$\eta$ as
%%%%%%%%%%%%%%%%%%%%%%%%%%%%%%%%%%%%%%%%%%%%%%%%%%%%%%%%%%%%%%%%%%%
\begin{equation}
  ds^2 = a^2(\eta) \left[-d\eta^2+dr^2+r^2 d\Omega^2 \right] .
\label{metric}
\end{equation}
%%%%%%%%%%%%%%%%%%%%%%%%%%%%%%%%%%%%%%%%%%%%%%%%%%%%%%%%%%%%%%%%%%%
Since our fiducial observer is located at the origin of the
coordinates ($\eta=\eta_0$, $r=0$), an object at $r$ and $\eta$ on the
light-cone hypersurface satisfies the simple relation
$r=\eta_0-\eta$.

We denote the comoving number density of observed objects (galaxies or
QSOs satisfying the selection criteria) at $\eta$ and $\bfx=(r,\bft)$
by $n(\eta,\bfx)$. Then the corresponding number density defined on
the light-cone is written as
%%%%%%%%%%%%%%%%%%%%%%%%%%%%%%%%%%%%%%%%%%%%%%%%%%%%%%%%%%%%%%%%%%%
\begin{equation}
  \nLC(r,\bft)=n(\eta_0-r,r,\bft) .
\label{eq:nlc1}
\end{equation}
%%%%%%%%%%%%%%%%%%%%%%%%%%%%%%%%%%%%%%%%%%%%%%%%%%%%%%%%%%%%%%%%%%%
If we introduce the mean {\it observed} number density (comoving) and
the density fluctuation at $\eta$, $n_0(\eta)$ and
$\Delta(\eta,\bfx)$, on the constant-time hypersurface,
%%%%%%%%%%%%%%%%%%%%%%%%%%%%%%%%%%%%%%%%%%%%%%%%%%%%%%%%%%%%%%%%%%%
\begin{equation}
  n(\eta,\bfx) = n_0(\eta) \left[1+\Delta(\eta,\bfx)\right],
\end{equation}
%%%%%%%%%%%%%%%%%%%%%%%%%%%%%%%%%%%%%%%%%%%%%%%%%%%%%%%%%%%%%%%%%%%
Eq. (\ref{eq:nlc1}) can be rewritten as
%%%%%%%%%%%%%%%%%%%%%%%%%%%%%%%%%%%%%%%%%%%%%%%%%%%%%%%%%%%%%%%%%%%
\begin{equation}
  \nLC(r,\bft)=n_0(\eta_0-r) \left[1+\Delta(\eta_0-r,r,\bft) \right] .
\label{nLC}
\end{equation}
%%%%%%%%%%%%%%%%%%%%%%%%%%%%%%%%%%%%%%%%%%%%%%%%%%%%%%%%%%%%%%%%%%%
The {\it observed} number density $n_0(\eta)$ is different from the true
density of the objects $\overline{n}(\eta)$ at $\eta$ by a factor of the
selection function $\phi(\eta)$:
%%%%%%%%%%%%%%%%%%%%%%%%%%%%%%%%%%%%%%%%%%%%%%%%%%%%%%%%%%%%%%%%%%%
\begin{equation}
  n_0(\eta) = \overline{n}(\eta) \phi(\eta).
\label{eq:n0}
\end{equation}
%%%%%%%%%%%%%%%%%%%%%%%%%%%%%%%%%%%%%%%%%%%%%%%%%%%%%%%%%%%%%%%%%%%
Thus $n_0(\eta)$ already includes the selection criteria, which depend on
the luminosity function of the objects and thus the magnitude-limit of
the survey, for instance:

When $\nLC(r,\bft)$ is given, one may compute the following two-point
statistics:
%%%%%%%%%%%%%%%%%%%%%%%%%%%%%%%%%%%%%%%%%%%%%%%%%%%%%%%%%%%%%%%%%%%
\begin{eqnarray}
  {\cal X}(R) &=& {1\over V^{\rm LC}}
  \int{d\Omega_{\hat \bfR}\over 4\pi} 
  \int  r_1^2 dr_1 d\Omega_{\bft_1} 
  \int  r_2^2 dr_2 d\Omega_{\bft_2} 
\nonumber \\
&\times&  \nLC(r_1,\bft_1)\nLC(r_2,\bft_2)
  \delta^{(3)}(\bfx_1-\bfx_2-\bfR).
\label{C3}
\end{eqnarray}
%%%%%%%%%%%%%%%%%%%%%%%%%%%%%%%%%%%%%%%%%%%%%%%%%%%%%%%%%%%%%%%%%%%
Here $\bfx_1=(r_1,r_1\bft_1)$, $\bfx_2=(r_2,r_2\bft_2)$, 
$R=|\bfR|$, $\hat \bfR=\bfR/R$, and $V^{\rm LC}$ is the comoving
survey volume of the data catalogue:
%%%%%%%%%%%%%%%%%%%%%%%%%%%%%%%%%%%%%%%%%%%%%%%%%%%%%%%%%%%%%%%%%%%%%%%%%%%%
\begin{equation}
  V^{\rm LC} =\int_\rmin^\rmax  r^2 dr 
    \int d\Omega_\bft={4\pi \over 3} (\rmax^3 - \rmin^3) ,
\end{equation}
%%%%%%%%%%%%%%%%%%%%%%%%%%%%%%%%%%%%%%%%%%%%%%%%%%%%%%%%%%%%%%%%%%%%%%%%%%%%
with $\rmax = r(z_{\rm max})$ and $\rmin= r(z_{\rm min})$ the
boundaries of the survey volume. Although the second equality as well
as the analysis below assumes that the survey volume extends to $4\pi$
steradian, all the results below can be easily generalized to the case
of the finite angular extent.

Substituting Eq. (\ref{nLC}), the ensemble average of an
estimator ${\cal X}(R)$ can be explicitly written as
%%%%%%%%%%%%%%%%%%%%%%%%%%%%%%%%%%%%%%%%%%%%%%%%%%%%%%%%%%%%%%%%%%%
\begin{equation}
\bigl<{\cal X}(R)\bigr>={\cal W}(R)+{\cal U}(R),
\end{equation}
%%%%%%%%%%%%%%%%%%%%%%%%%%%%%%%%%%%%%%%%%%%%%%%%%%%%%%%%%%%%%%%%%%%
where
%%%%%%%%%%%%%%%%%%%%%%%%%%%%%%%%%%%%%%%%%%%%%%%%%%%%%%%%%%%%%%%%%%%
\begin{eqnarray}
  &&{\cal W}(R)=
  {1\over V^{\rm LC}}
  \int{d\Omega_{\hat \bfR}\over 4\pi} 
  \int dr_1 r_1^2 \int d\Omega_{\bft_1} 
  \int dr_2 r_2^2 \int d\Omega_{\bft_2}
  n_0(\eta_0-r_1)n_0(\eta_0-r_2)
\nonumber
\\
  &&\hspace{1.5cm}
  \times
  \Bigl<\Delta(\eta_0-r_1,r_1,\bft_1)\Delta(\eta_0-r_2,r_2,\bft_2)\Bigr>
  \delta^{(3)}(\bfx_1-\bfx_2-\bfR) ,
\label{C4}
\end{eqnarray}
%%%%%%%%%%%%%%%%%%%%%%%%%%%%%%%%%%%%%%%%%%%%%%%%%%%%%%%%%%%%%%%
and
%%%%%%%%%%%%%%%%%%%%%%%%%%%%%%%%%%%%%%%%%%%%%%%%%%%%%%%%%%%%%%%%%%%
\begin{eqnarray}
  &&{\cal U}(R)=
  {1\over V^{\rm LC}}
  \int{d\Omega_{\hat \bfR}\over 4\pi} 
  \int dr_1 r_1^2 \int d\Omega_{\bft_1} 
  \int dr_2 r_2^2 \int d\Omega_{\bft_2}
\nonumber
\\
  &&\hspace{1.5cm}
  \times
  n_0(\eta_0-r_1)n_0(\eta_0-r_2)
  \delta^{(3)}(\bfx_1-\bfx_2-\bfR) .
\label{C33}
\end{eqnarray}
%%%%%%%%%%%%%%%%%%%%%%%%%%%%%%%%%%%%%%%%%%%%%%%%%%%%%%%%%%%%%%%%%%%
After a tedious but straightforward calculation,\cite{YS} we have shown that
the above definitions can be approximated as
%%%%%%%%%%%%%%%%%%%%%%%%%%%%%%%%%%%%%%%%%%%%%%%%%%%%%%%%%%%%%%%%%%%
\begin{subeqnarray}
  {\cal W}(R) &\simeq& {4\pi\over V^{\rm LC }} 
  \int_\rmin^\rmax r^2 dr \left[n_0(\eta_0-r)\right]^2 
                    \xi(R;\eta_0-r)_{\rm Source}~,
  \slabel{B69} \\
  {\cal U}(R) &\simeq&
  {4\pi\over V^{\rm LC }} \int_\rmin^{\rmax}   r^2 dr
   \left[n_0(\eta_0-r)\right]^2 ,
  \slabel{C81}
\end{subeqnarray}
%%%%%%%%%%%%%%%%%%%%%%%%%%%%%%%%%%%%%%%%%%%%%%%%%%%%%%%%%%%%%%%%%%%
where $\xi(R;\eta)_{\rm Source}$ is the conventional two-point
correlation defined on the constant hypersurface at the source's
position. Note that ${\cal U}(R)$ is independent of $R$ for $R\ll\rmax$,
as expected.

The next task is to define the two-point correlation function on the
light-cone. We propose the definition
%%%%%%%%%%%%%%%%%%%%%%%%%%%%%%%%%%%%%%%%%%%%%%%%%%%%%%%%%%%%%%%%%%%
\begin{equation}
  \xi^{\rm LC}(R) 
\equiv {{\cal W}(R)\over {\cal U}(R)}
={\displaystyle {\int_\rmin^\rmax dr r^2 
  n_0(\eta_0-r)^2 \xi(R;\eta_0-r)_{\rm Source}}
  \over
  {\displaystyle \int_\rmin^\rmax dr r^2 
  n_0(\eta_0-r){}^2}}~,
\label{eq:xiA}
\end{equation}
%%%%%%%%%%%%%%%%%%%%%%%%%%%%%%%%%%%%%%%%%%%%%%%%%%%%%%%%%%%%%%%%%%%
where the second expression uses Eqs. (\ref{B69}) and (\ref{C81}).
If the correlation function of objects does not evolve, i.e.,
$\xi(R;\eta_0-r)_{\rm Source} = \xi(R;\eta_0)_{\rm Source}$, Eq.
(\ref{eq:xiA}) readily yields
%%%%%%%%%%%%%%%%%%%%%%%%%%%%%%%%%%%%%%%%%%%%%%%%%%%%%%%%%%%%%%%%%%%
\begin{equation}
\xi^{\rm LC}(R) = \xi(R;\eta_0)_{\rm Source},
\label{eq:xiAS}
\end{equation}
%%%%%%%%%%%%%%%%%%%%%%%%%%%%%%%%%%%%%%%%%%%%%%%%%%%%%%%%%%%%%%%%%%%
as should be the case. 

Equation (\ref{eq:xiA}) can be directly evaluated from any observed
sample. First, average over the angular distribution and estimate the
differential redshift number count $dN/dz$ of the objects. Second,
distribute random particles over the whole sample volume so that they
obey the same $dN/dz$. Then the conventional pair-count between the
objects and random particles yields ${\cal X}(R)$ (although not
$\bigl<{\cal X}(R)\bigr>$, of course), while ${\cal U}(R)$ can be
estimated from the pair-count of the random particles themselves.

\subsection{Predicting two-point correlation functions on a light-cone
\label{subsec:predictxi}}

The corresponding theoretical predictions can be easily computed also,
once a set of cosmological parameters and a model for the evolution of
bias are specified. To illustrate the behavior of the two-point
correlation functions on the light-cone, we adopt the following
models.

\medskip

%%%%%%%%%%%%%%%%%%%%%%%%%%%%%%%%%%%%%%%%%%%%%%%%%%%%%%%%%%%%%%%%%%%%%%
\begin{description}
\item[(i) cosmological parameters]: We consider three models based on
  cold dark matter (CDM) cosmogonies; SCDM with
  $(\Omega_0,\lambda_0,h,\sigma_8)=(1.0,0.0,0.5,0.56)$, and LCDM with
  $(\Omega_0,\lambda_0,h,\sigma_8)=(0.3,0.7,0.7,1.0)$. The
  normalization $\sigma_8$ is determined from the cluster
  abundances.\cite{KS} Then we use the following fitting formula\cite{BBKS} for
  the linear power-spectrum of mass fluctuation:
%%%%%%%%%%%%%%%%%%%%%%%%%%%%%%%%%%%%%%%%%%%%%%%%%%%%%%%%%%%%%%%
\begin{subeqnarray}
\slabel{eq:cdmspect}
  \Delta^2_\L (k_\L) &\propto& 
{q^4[{\ln(1+2.34q)/2.34 q}]^2
\over 
\sqrt{1 + 3.89 q + (16.1q)^2 + (5.46q)^3 + (6.71q)^4}},
\\
q &\equiv& k_\L/(\Gamma h {\rm Mpc}^{-1}) , \\
\slabel{eq:gamma}
  \Gamma &\equiv& \Omega_0 h (T_{\gamma 0}/2.7 \mbox{ K})^{-2}
    \exp[-\Omega_{\rm b}(1+\sqrt{2 h}\Omega_0^{-1})] .
\end{subeqnarray}
%%%%%%%%%%%%%%%%%%%%%%%%%%%%%%%%%%%%%%%%%%%%%%%%%%%%%%%%%%%%%%%

\item[(ii) mass correlation function]: Gravitational nonlinear
  evolution is included by using the following fitting formulae\cite{PD94,PD96}
  for the mass power spectrum:
%%%%%%%%%%%%%%%%%%%%%%%%%%%%%%%%%%%%%%%%%%%%%%%%%%%%%%%%%%%%%%%%%%%%%
\begin{subeqnarray}
\label{eq:pdfit}
k_\L &=& {k_\NL \over 
\left[1+\Delta^2_\NL(k_\NL)\right]^{1/3}}, \\
\Delta^2_\NL(k_\NL) &=& 
  \tilde f_\NL[\Delta^2_\L(k_\L)] , \\
  \tilde f_\NL(x) &=& x \left[ {1+B\beta x+(Ax)^{\alpha\beta}
\over 1 + [(Ax)^\alpha g^3(z)/(V\sqrt{x})]^\beta
}\right] , \\
A &=& 0.482 (1+n/3)^{-0.947} ,\\
B &=& 0.226 (1+n/3)^{-1.778} ,\\
\alpha &=& 3.310 (1+n/3)^{-0.244} ,\\
\beta &=& 0.862 (1+n/3)^{-0.287} ,\\
V &=& 11.55 (1+n/3)^{-0.423} ,\\
n_{\rm eff}(k) &=& \left| {d \ln \Delta^2_\L(k) \over d \ln k}
 \right|_{k=k_\L} -3 .
\end{subeqnarray}
%%%%%%%%%%%%%%%%%%%%%%%%%%%%%%%%%%%%%%%%%%%%%%%%%%%%%%%%%%%%%%%%%%%%%
Then the nonlinear mass correlation function is computed via
%%%%%%%%%%%%%%%%%%%%%%%%%%%%%%%%%%%%%%%%%%%%%%%%%%%%%%%%%%%%%%
\begin{equation}
\xi_\NL (x) = 
\int_0^\infty \Delta^2(k) { {\rm sin} kx \over x} {dk \over k} .
\end{equation}
%%%%%%%%%%%%%%%%%%%%%%%%%%%%%%%%%%%%%%%%%%%%%%%%%%%%%%%%%%%%%%

\item[(iii) evolution of bias]: This is by far the most uncertain
  factor in the current modeling. We simply use a linear bias
  model\cite{Fry96} on the basis of perturbation theory:
%%%%%%%%%%%%%%%%%%%%%%%%%%%%%%%%%%%%%%%%%%%%%%%%%%%%%%%%%%%%%%%%%%%
\begin{equation}
   b(z) = 1 + {D(0) \over D(z)} (b_0-1),
\label{eq:fryb}
\end{equation}
%%%%%%%%%%%%%%%%%%%%%%%%%%%%%%%%%%%%%%%%%%%%%%%%%%%%%%%%%%%%%%%%%%%
where $b_0$ is the present value of the bias parameter. We denote by
$D(z)$ the linear growth rate (normalized as $D(z)=1/(1+z)$
for $z\rightarrow \infty$):
%%%%%%%%%%%%%%%%%%%%%%%%%%%%%%%%%%%%%%%%%%%%%%%%%%%%%%%%%%%%%%%%%%%
\begin{equation}
   D(z) = {5\Omega_0H_0^2 \over2}H(z)
   \int_z^\infty {1+z' \over H(z')^3}\, dz'.
\end{equation}
%%%%%%%%%%%%%%%%%%%%%%%%%%%%%%%%%%%%%%%%%%%%%%%%%%%%%%%%%%%%%%%%%%%
Here $H(z)$ is the Hubble parameter at redshift $z$:
%%%%%%%%%%%%%%%%%%%%%%%%%%%%%%%%%%%%%%%%%%%%%%%%%%%%%%%%%%%%%%%%%%%
\begin{equation}
   H(z) = H_0\sqrt{\Omega_0 (1 + z)^3 +
(1-\Omega_0-\lambda_0) (1 + z)^2 + \lambda_0} .
\end{equation}
%%%%%%%%%%%%%%%%%%%%%%%%%%%%%%%%%%%%%%%%%%%%%%%%%%%%%%%%%%%%%%%%%%%
According to this simplified scheme, $\xi(x;z)_{\rm Source}$ is given by
$b(z)\xi_\NL (x;z)$.

\item[(iv) selection function]: With a magnitude-limited QSO sample in
  mind, we adopt the following B-band quasar luminosity
  function.\cite{WN,NS} For $0.3<z<3$,
%%%%%%%%%%%%%%%%%%%%%%%%%%%%%%%%%%%%%%%%%%%%%%%%%%%%%%%%%%%%%%%%%%%
\begin{subeqnarray} 
\Phi(M_{\rm B},z) &=& {\Phi_* \over
10^{0.4(\alpha+1)[M_{\rm B}-M_{\rm B}^*(z)]} + 10^{0.4(\beta+1)[M_{\rm
B}-M_{\rm B}^*(z)]} } , \label{eq:phiqso} \\ 
M_{\rm B}^*(z) &=& M_{\rm B}^* - 2.5 \kappa_\L \log(1+z) , 
\end{subeqnarray}
%%%%%%%%%%%%%%%%%%%%%%%%%%%%%%%%%%%%%%%%%%%%%%%%%%%%%%%%%%%%%%%%%%%
with $M_{\rm B}^*= -20.91+5\log h$, $\kappa_\L =3.15$, $\alpha=-3.79$,
$\beta=-1.44$, $\Phi_* = 6.4\times 10^{-6} h^3{\rm Mpc}^3$. For $z>3$,
%%%%%%%%%%%%%%%%%%%%%%%%%%%%%%%%%%%%%%%%%%%%%%%%%%%%%%%%%%%%%%%%%%%
\begin{subeqnarray} 
\Phi(M_{\rm B},z) &=& {\Phi_* \times
10^{-[A+0.4B(\beta+1)]} \over 10^{0.4(\alpha+1)[M_{\rm B}-M_{\rm
B}^*(z)]} + 10^{0.4(\beta+1)[M_{\rm B}-M_{\rm B}^*(z)]} } ,
\label{eq:phiqso3} 
\\ M_{\rm B}^*(z) &=& M_{\rm B}^* - 2.5 \kappa_\L  \log 4 + B, 
\end{subeqnarray}
%%%%%%%%%%%%%%%%%%%%%%%%%%%%%%%%%%%%%%%%%%%%%%%%%%%%%%%%%%%%%%%%%%%
with $A=(z-3)\log 3.2$, $B=2.5A/(\alpha-\beta)$.  To compute the
B-band apparent magnitude from a quasar of absolute magnitude $M_{\rm
  B}$ at $z$ (with the luminosity distance $d_\L$), we apply the
K-correction,
%%%%%%%%%%%%%%%%%%%%%%%%%%%%%%%%%%%%%%%%%%%%%%%%%%%%%%%%%%%%%%%%%%%
\begin{equation}
  B = M_{\rm B} + 5 \log(d_{\rm L}(z)/ 10 {\rm pc}) 
- 2.5(1-p)\log (1+z), 
\end{equation}
%%%%%%%%%%%%%%%%%%%%%%%%%%%%%%%%%%%%%%%%%%%%%%%%%%%%%%%%%%%%%%%%%%%
for the quasar energy spectrum $L_\nu \propto \nu^{-p}$ with $p=0.5$.
While this luminosity function is derived from observed data assuming
$\Omega_0$ and $\lambda_0$, we use this also for other cosmological
models to make clean the differences due purely to the light-cone effect.
\end{description}
%%%%%%%%%%%%%%%%%%%%%%%%%%%%%%%%%%%%%%%%%%%%%%%%%%%%%%%%%%%%%%%%%%%

%%%%%%%%%%%%%%%%%%%%%%%%%%%%%%%%%%%%%%%%%%%%%%%%%%%%%%%%%%%%%%%%%%%%%
\begin{figure}[tbh]
\begin{center}
    \leavevmode\epsfysize=6cm \epsfbox{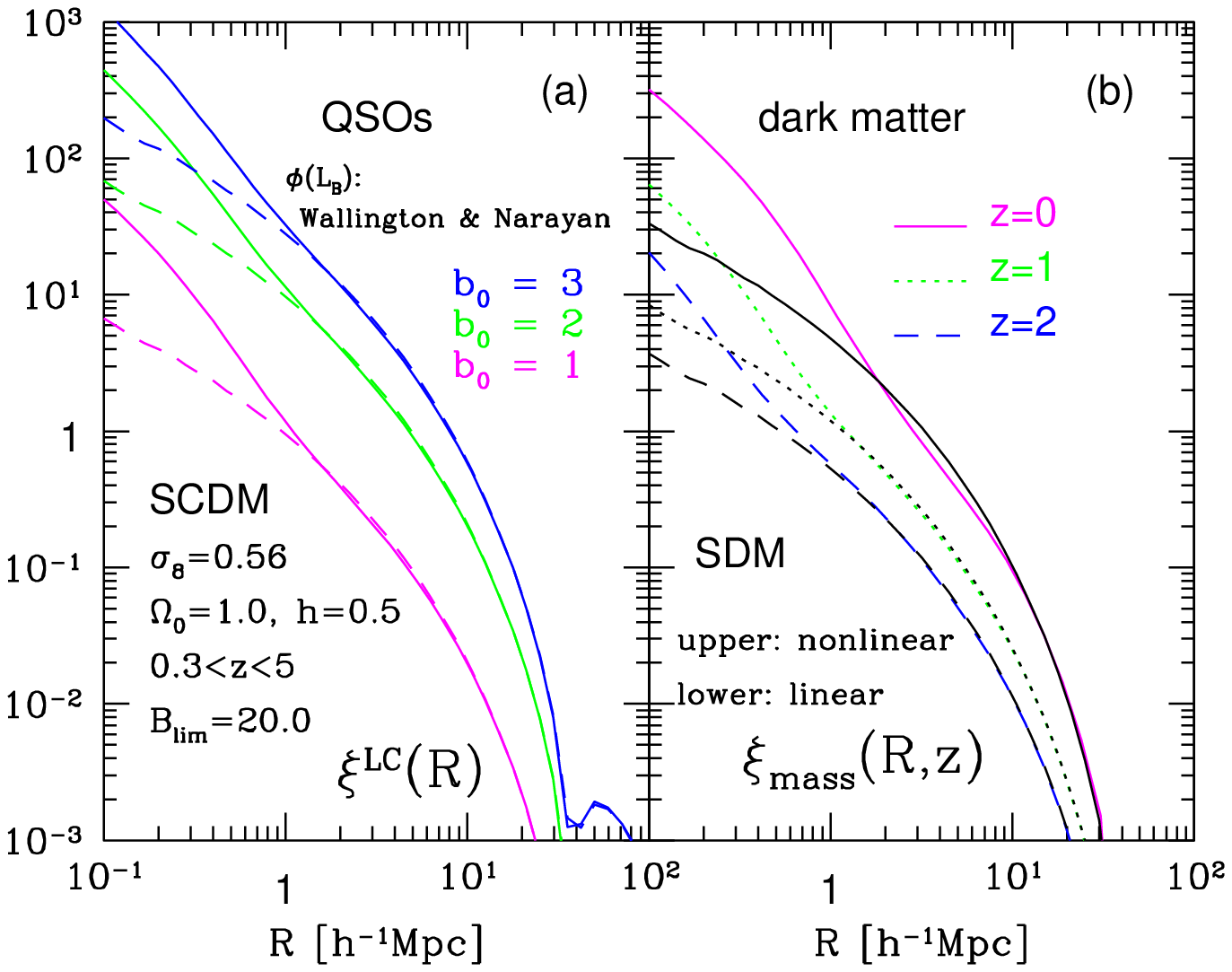}
\end{center}
\caption{  Two-point correlation functions in the cluster normalized 
  standard CDM model.  (a) $\xi^{LC}(R)$ defined on the light-cone
  hypersurface for QSOs with $B_{lim}=20$. We assume three cases for
  the biasing parameter, $b_0=3$, 2 and 1, from top to bottom.
  A nonlinear mass correlation function\protect\cite{PD96}\protect is
  used for the solid lines, while the linear theory is used for dashed lines.
  (b) Linear (lower curves) and nonlinear\protect\cite{PD96}\protect
  (upper curves) mass correlation functions defined on the constant-time
  hypersurfaces $z=0$, 1 and 2.
  \label{fig:lcxiR_scdm_fry}}
\vspace*{0.5cm}
\begin{center}
    \leavevmode\epsfysize=6cm \epsfbox{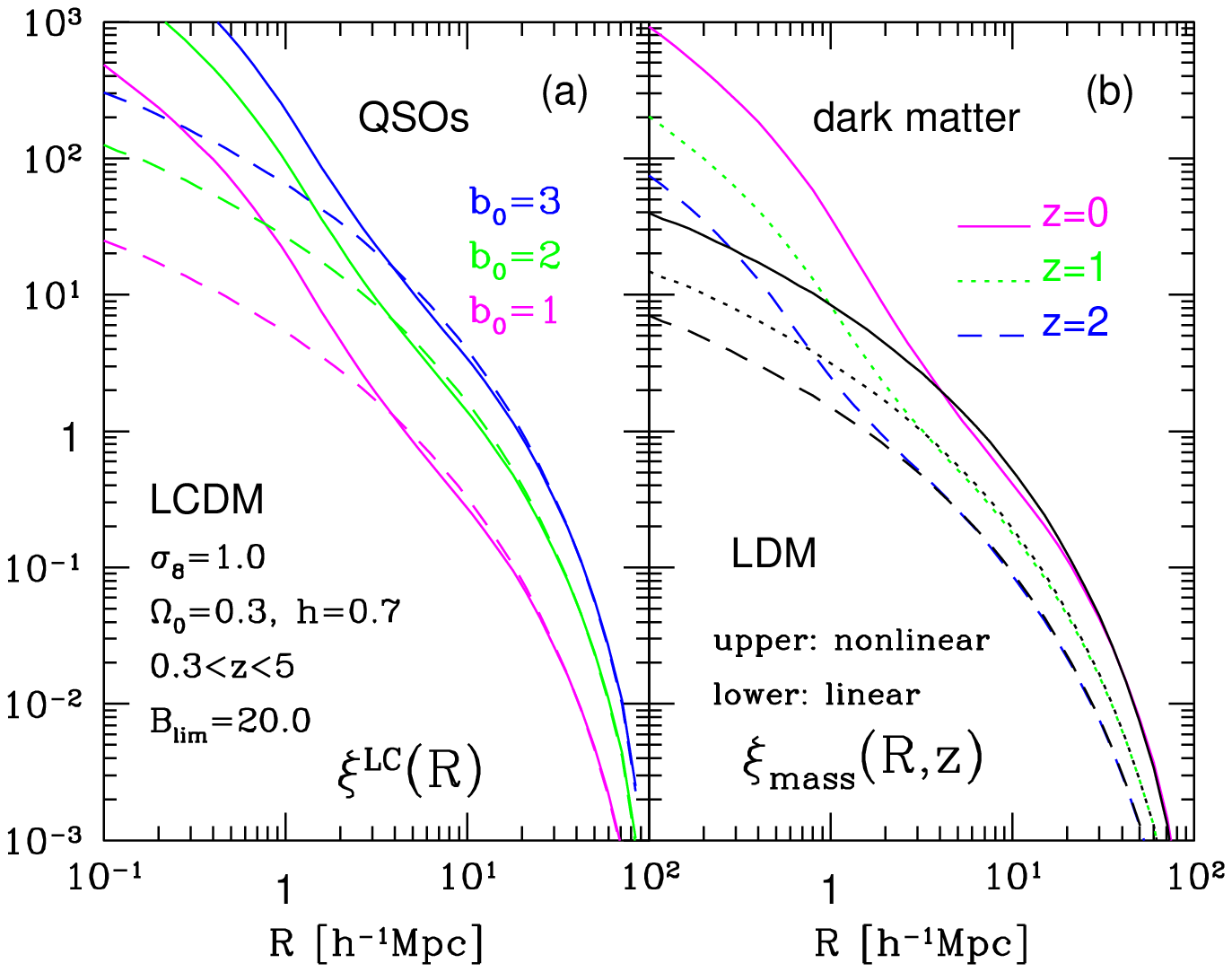}
\end{center}
\caption{ Same as Fig.~\protect\ref{fig:lcxiR_scdm_fry}\protect, but in
  a cluster normalized, low-density spatially flat CDM model.
  \label{fig:lcxiR_lcdm_fry}}
\end{figure}
%%%%%%%%%%%%%%%%%%%%%%%%%%%%%%%%%%%%%%%%%%%%%%%%%%%%%%%%%%%%%%%%%%%%%

\begin{figure}[tbh]
\begin{center}
    \leavevmode\epsfysize=6cm \epsfbox{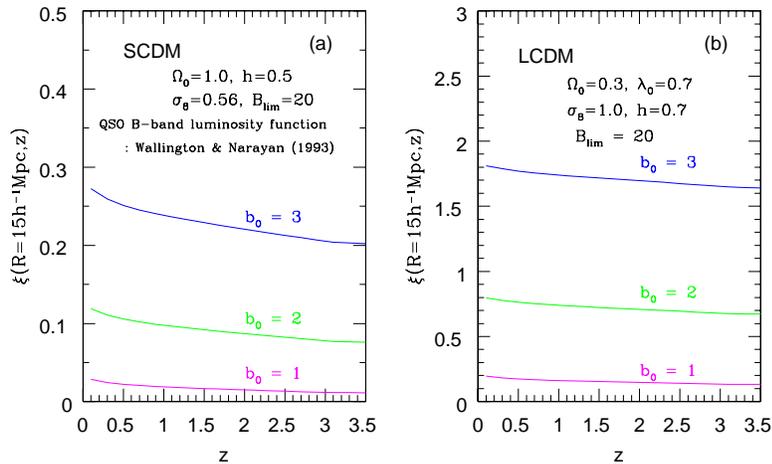}
\end{center}
\caption{ Evolution of amplitudes of two-point correlation functions at
$R=15\himpc$ of QSOs on the light-cone in cluster normalized
(a) standard and (b) low-density spatially flat CDM models. 
  \label{fig:lcxiz_fry}}
\end{figure}

The results are plotted in Figs.~\ref{fig:lcxiR_scdm_fry} and
\ref{fig:lcxiR_lcdm_fry} for SCDM and LCDM models, respectively. The
B-band limiting magnitude $B_{lim}=20$ roughly corresponds to the
upcoming SDSS QSO sample. The corresponding evolution of the amplitude
of the correlation at $R=15$\himpc is plotted in
Fig.~\ref{fig:lcxiz_fry}. In the specific bias model we adopted, the
amplitude monotonically decreases with increasing $z$. This is
inconsistent with an observational claim\cite{FAC} that the QSO
correlation amplitude {\it increases} as $z$.  Given the theoretical
uncertainties of the current theoretical understanding of the bias, it
is premature to draw any decisive conclusion at this point. In fact,
behavior consistent with the observational claim is obtained with a
different model of bias.\cite{YS,Mataresse} Nevertheless, this example
illustrates the potential importance of the light-cone effect in
understanding the evolution of the bias of high-redshift objects.

%%%%%%%%%%%%%%%%%%%%%%%%%%%%%%%%%%%%%%%%%%%%%%%%%%%%%%%%%%%%%%%%%%%%%
%%%%%%%%%%%%%%%%%%%%%%%%%%%%%%%%%%%%%%%%%%%%%%%%%%%%%%%%%%%%%%%%%%%%%

\subsection{Higher-order statistics on light-cone \label{subsec:highlc}}

Let us move to the higher-order statistics of clustering on the
light-cone hypersurface. In particular we focus on the volume-averaged
$N$-point correlation functions, $\overline{\xi}_N(R;z)$, at a redshift
$z$ and on a comoving smoothing scale $R$.  In the higher-order
statistics, it is more useful to introduce the normalized higher-order
moments $S_N(R;z) \equiv
\overline{\xi}_N(R;z)/[\overline{\xi}_2(R;z)]^{N-1}$ than
$\overline{\xi}_N(R;z)$.  The hierarchical clustering ansatz states that
$S_N(R;z)$ is constant and independent of the scale $R$.  Moreover,
$\overline{\xi}_N(R;z)$ evolves in proportion to
$\left[\overline{\xi}_2(R;z)\right]^{N-1}$, and $S_N(R;z)$ is
independent of $z$, according to perturbation theory.\cite{Fry84,Bouchet}

As described in \S \ref{subsec:defxi} and \S \ref{subsec:predictxi},
however, the $N$-point correlation functions averaged over the
light-cone,
%%%%%%%%%%%%%%%%%%%%%%%%%%%%%%%%%%%%%%%%%%%%%%%%%%%%%%%%%%%%%%%%%%%%%
\begin{equation}
  \overline{\xi}_N(R;<\zm) \equiv {\displaystyle 
\int_0^{\zm} z^2 dz ~w(z) \overline{\xi}_N(R;z) \,  \over
\displaystyle \int_0^{\zm} z^2 dz ~w(z)} ,
  \label{eq:xinzm}
\end{equation}
%%%%%%%%%%%%%%%%%%%%%%%%%%%%%%%%%%%%%%%%%%%%%%%%%%%%%%%%%%%%%%%%%%%%%
and the corresponding moments,
%%%%%%%%%%%%%%%%%%%%%%%%%%%%%%%%%%%%%%%%%%%%%%%%%%%%%%%%%%%%%%%%%%%%%
\begin{equation}
 \overline{S_N}(R;<\zm) \equiv
{\overline{\xi}_N(R;<\zm)
\over \left[\overline{\xi}_2(R;<\zm)\right]^{N-1}},  
  \label{eq:snzm}
\end{equation}
%%%%%%%%%%%%%%%%%%%%%%%%%%%%%%%%%%%%%%%%%%%%%%%%%%%%%%%%%%%%%%%%%%%%%
are the statistics more directly estimated from redshift surveys than
their counterparts defined on the idealistic $z=0$ hypersurface.  Note
that the expression (\ref{eq:xinzm}) looks slightly different from
Eq. (\ref{eq:xiA}). This is because we have count-in-cell analysis
in mind with the sampling cells being placed randomly in the $z$-coordinate.
The effect of the selection function is taken into account by the weight
function $w(z)$. In the case of count-in-cell analysis, one can correct
for the selection function $\phi(z)$ by multiplying the count in cells
located at $z$ by $1/\phi(z)$. Then $w(z)$ can be set to unity for
$z<\zm$ and zero for $z>\zm$ in principle, where $\zm$ is the maximum
redshift of the sample. In the remainder of this subsection, we assume that
the effect of the selection function is already corrected in this way,
and consider the light-cone effect on the moments (\ref{eq:snzm}) due to
the difference of the gravitational evolution within the survey volume.

Let us define a function $G$ which describes the evolution of the averaged
two-point correlation function at $R$ and $z$:
%%%%%%%%%%%%%%%%%%%%%%%%%%%%%%%%%%%%%%%%%%%%%%%%%%%%%%%%%%%%%%%%%%%%%
\begin{equation}
  \overline{\xi}(R;z) = G(R;z) \overline{\xi}(R;0) .
  \label{eq:gz}
\end{equation}
%%%%%%%%%%%%%%%%%%%%%%%%%%%%%%%%%%%%%%%%%%%%%%%%%%%%%%%%%%%%%%%%%%%%%
In general, $G$ is not a simple function of $R$ and $z$, but a
complicated functional of $\xi$. In the linear regime, however, $G$ is
independent of $R$ and given by $[D(z)/D(0)]^2$, and even in the
nonlinear regime, it is known that $G$ is approximately expressed as a
function of $R$ and $z$ alone. To proceed more specifically, we apply
the fitting formula\cite{JMW} which relates the evolved two-point
correlation function $\overline{\xi}_E(R;z)$ with its linear
counterpart $\overline{\xi}_L(R_0;z)$ as follows:
%%%%%%%%%%%%%%%%%%%%%%%%%%%%%%%%%%%%%%%%%%%%%%%%%%%%%%%%%%%%%%%%%%%%%
\begin{subeqnarray}
  \slabel{eq:jmwfit1}
\overline{\xi}_E(R;z) &=& B(n) \, F[\overline{\xi}_L(R_0;z)/B(n)] ,\\
  \slabel{eq:jmwfit4}
 F(x) &=& {x+0.45 x^2 -0.02 x^5+0.05 x^6 \over 1+0.02 x^3 + 0.003 x^{9/2}} .
\end{subeqnarray}
%%%%%%%%%%%%%%%%%%%%%%%%%%%%%%%%%%%%%%%%%%%%%%%%%%%%%%%%%%%%%%%%%%%%%
In the above equations, $n$ denotes the effective spectral index of
the power spectrum evaluated at the scale just entering the nonlinear
regime, $R_0 =[1 + \overline{\xi}_E(z,R)]^{1/3} R$, and $B(n) =
[(3+n)/3]^{0.8}$.

The inverse of Eq. (\ref{eq:jmwfit1}) is also
empirically fitted as follows:
%%%%%%%%%%%%%%%%%%%%%%%%%%%%%%%%%%%%%%%%%%%%%%%%%%%%%%%%%%%%%%%%%%%%%
\begin{subeqnarray}
\overline{\xi}_L(R_0;z) &=& B(n) F^{-1}[\overline{\xi}_E(R;z)/B(n)] ,\\
F^{-1}(y) &=& y\left( {1+ 0.036 y^{1.93} +0.0001 y^3
       \over   1+1.75 y -0.0015 y^{3.63} +0.028 y^4}
      \right)^{1/3} .
  \slabel{eq:jmwinv2}
\end{subeqnarray}
%%%%%%%%%%%%%%%%%%%%%%%%%%%%%%%%%%%%%%%%%%%%%%%%%%%%%%%%%%%%%%%%%%%%%
Then the scale-dependent evolution factor $G(z)=G(R;z)$ defined by
 Eq. (\ref{eq:gz}) is expressed explicitly in terms of
 $\overline{\xi}_E(R;0)$:
%%%%%%%%%%%%%%%%%%%%%%%%%%%%%%%%%%%%%%%%%%%%%%%%%%%%%%%%%%%%%%%%%%%%%
\begin{equation}
G(R;z) \equiv  
{\overline{\xi}_E(R;z) \over \overline{\xi}_E(R;0)}
= \frac{B(n)}{\overline{\xi}_E(R;0)} F\left[{D^2(z) \over D^2(0)}  
       F^{-1}\left({\overline{\xi}_E(R,0) \over B(n)}\right) \right] .
  \label{eq:grz}
\end{equation}
%%%%%%%%%%%%%%%%%%%%%%%%%%%%%%%%%%%%%%%%%%%%%%%%%%%%%%%%%%%%%%%%%%%%%
Substituting the evolution law (\ref{eq:gz}), Eq. (\ref{eq:snzm})
is explicitly written as
%%%%%%%%%%%%%%%%%%%%%%%%%%%%%%%%%%%%%%%%%%%%%%%%%%%%%%%%%%%%%%%%%%%%%
\begin{eqnarray}
 \overline{S_N}(R;<\zm) &=& 
   \left[\int_0^{\zm} z^2 dz ~w(z)\right]^{N-2}
     \big/ \left[\int_0^{\zm} z^2 dz ~w(z) G(R;z)\right]^{N-1}
 \nonumber \\
&\times& \int_0^{\zm} z^2 dz~w(z) S_N(R;z) \left\{G(R;z)\right\}^{N-1}
\label{eq:m4}
\end{eqnarray}
%%%%%%%%%%%%%%%%%%%%%%%%%%%%%%%%%%%%%%%%%%%%%%%%%%%%%%%%%%%%%%%%%%%%%

\begin{figure}[tbh]
  \parbox{\halftext}{
\begin{center}
    \leavevmode\epsfysize=7cm \epsfbox{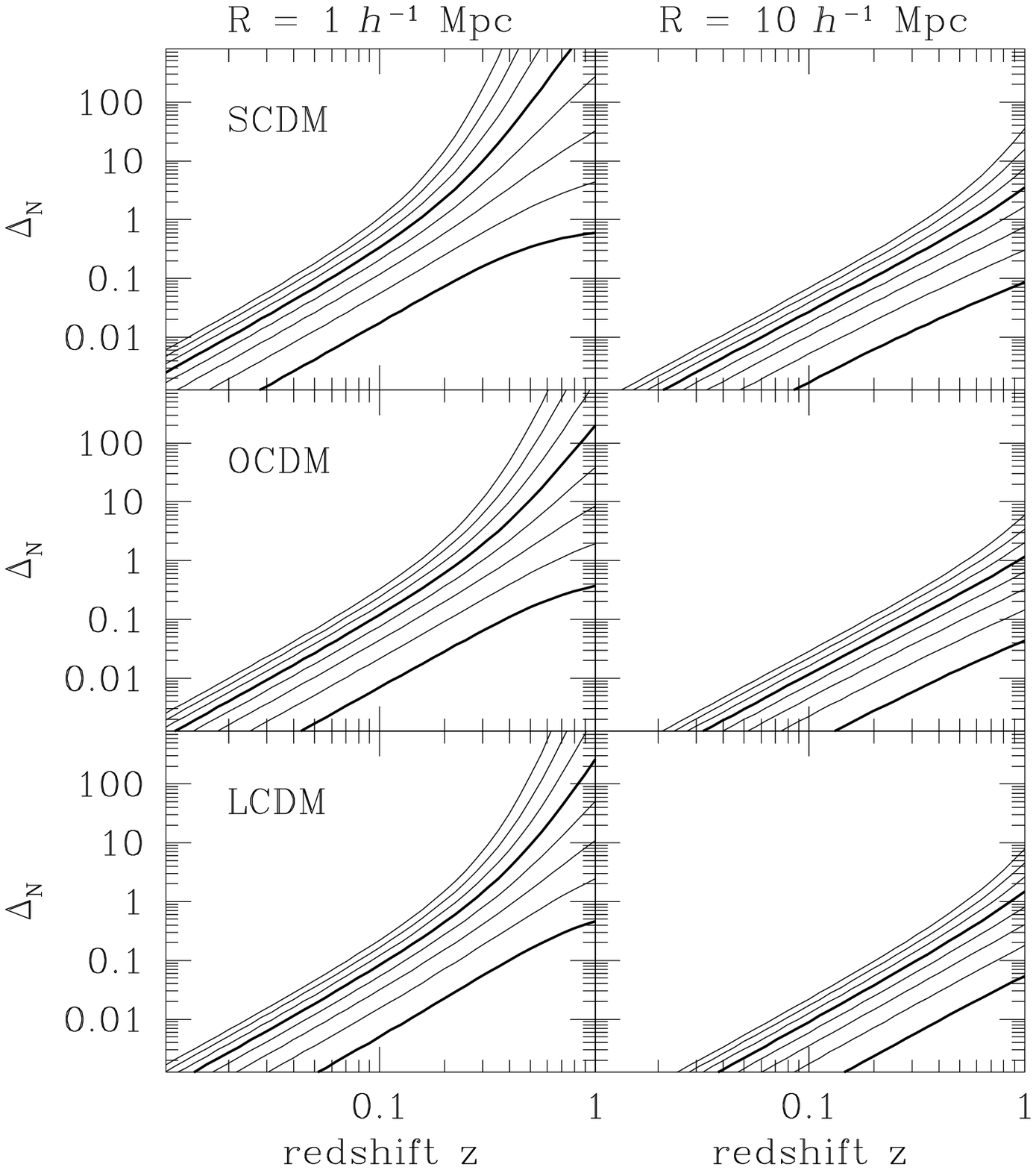}
\end{center}
\caption{ $\log_{10}\Delta_N(R;z)$ are shown as functions of
  $\log_{10}z$ at $R=1\himpc$ ({\it left panels}) and $10\himpc$ ({\it
  right panels}). The curves correspond to $N
  = 3,\ldots,N = 10$ from bottom to top with $N = 3$ and $N = 7$
  plotted in thick lines (from Ref.~\protect\citen{MSS}\protect).
  \label{fig:dnz}} } 
\hspace{0.1cm} 
\parbox{\halftext}{
\begin{center}
    \leavevmode\epsfysize=7cm \epsfbox{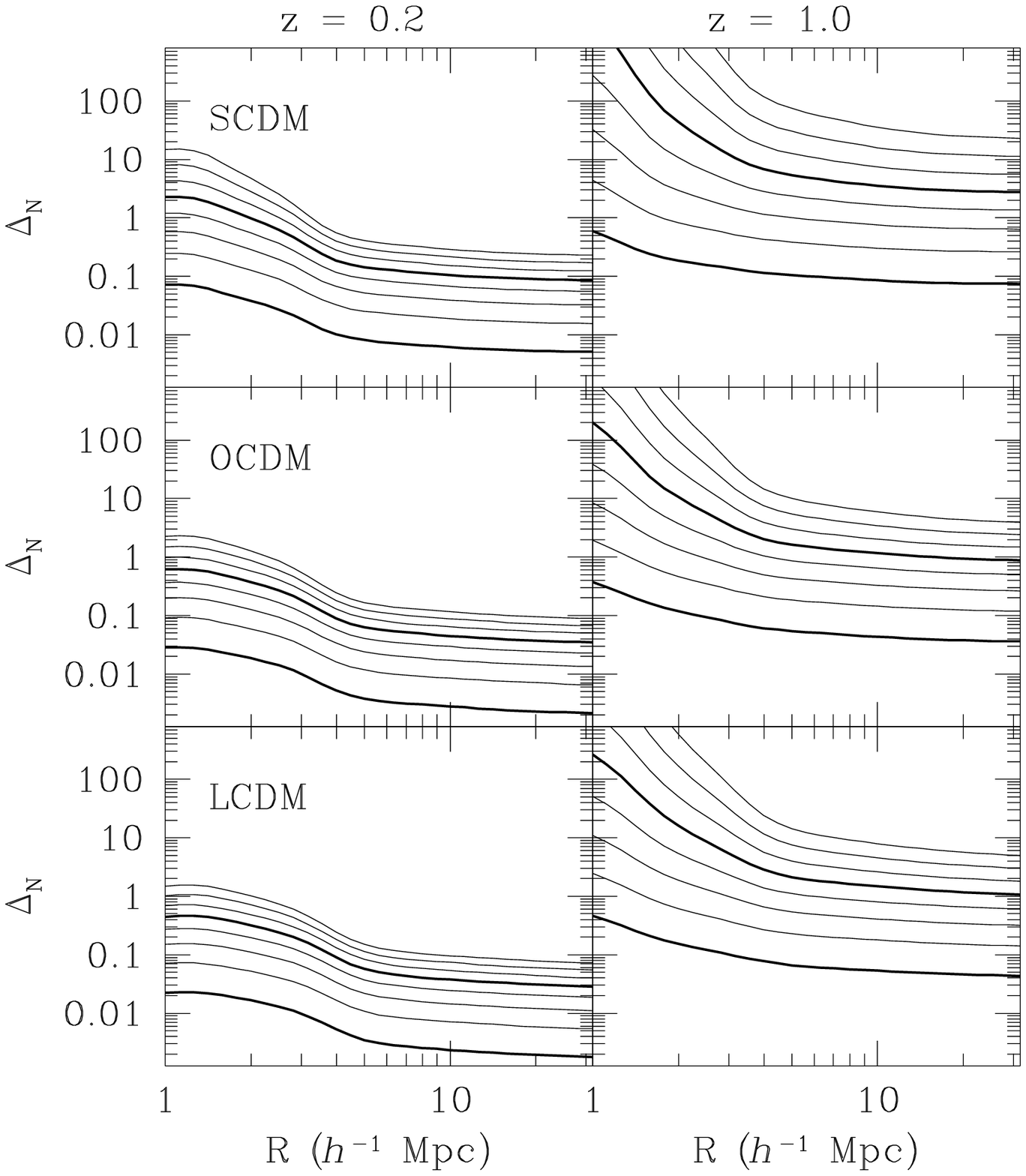}
\end{center}
\caption{$\log_{10}\Delta_N(R;z)$ are displayed as functions 
  of $\log_{10}R$ at $z=0.2$ ({\it left panels}) and $1.0$ ({\it right
    panels}). The curves correspond to $N
  = 3,\ldots,N = 10$ from bottom to top with $N = 3$ and $N = 7$
  plotted in thick lines \protect (from Ref.~\protect\citen{MSS}\protect).
\label{fig:dnr}}}
\end{figure}

In order to proceed further, we assume that $S_N(R;z)$ does not evolve
with $z$, i.e., $S_N(R;z) = S_N(R;0)$. As described above, this is a
reasonable approximation as long as objects are unbiased tracers of
the underlying density field.  Also let us introduce the measure of the
light-cone effect:
%%%%%%%%%%%%%%%%%%%%%%%%%%%%%%%%%%%%%%%%%%%%%%%%%%%%%%%%%%%%%%%%%%%%%
\begin{equation}
  \Delta_N(R;<\zm) \equiv {\overline{S_N}(R;<\zm) \over S_N(R;0)} - 1 .
  \label{eq:deltanz}
\end{equation}
%%%%%%%%%%%%%%%%%%%%%%%%%%%%%%%%%%%%%%%%%%%%%%%%%%%%%%%%%%%%%%%%%%%%%
Then $(1+\Delta_N)$ can be regarded as a correction factor for the
$S_N$ estimated {\em without} considering the light-cone effect.  This
quantifies the importance of the light cone effect on the higher-order
clustering statistics in future surveys.  Using Eqs.
(\ref{eq:m4}) and (\ref{eq:grz}) and assuming $S_N(z)= S_N(0)$, we
evaluate $\Delta_N(<\zm)$ for SCDM, LCDM, and OCDM, which have
$(\Omega_0,\lambda_0,h,\sigma_8)$ $=(1.0,0.0,0.5,0.56)$,
$(0.3,0.7,0.7,1.0)$, and $(0.3,0.7,0.0,1.0)$, respectively.
Figure~\ref{fig:dnz} displays $\Delta_N(R;<z)$ as a function of $z$,
while Figure~\ref{fig:dnr} plots $\Delta_N(R;<z)$ against $R$.

%%%%%%%%%%%%%%%%%%%%%%%%%%%%%%%%%%%%%%%%%%%%%%%%%%%%%%%%%%%%%%%%%%%%%
%%%%%%%%%%%%%%%%%%%%%%%%%%%%%%%%%%%%%%%%%%%%%%%%%%%%%%%%%%%%%%%%%%%%%

The figures suggest that the light-cone effect is quite
robust.Although its details depend on the model, the difference is
fairly small, and qualitatively all models behave similarly; the
magnitude of the correction monotonically increases for higher
order $N$. Also as expected, the light-cone effect becomes larger as
$\zm$ increases (Fig.~\ref{fig:dnz}). Although the correction is
relatively small for shallow surveys with $z \simlt 0.2$ samples,
$\Delta_N(R;<\zm)$ becomes $\simgt 10$\% in nonlinear scales ($R\sim
1\himpc$). In SCDM, for instance, $\Delta_N(R;<\zm)$ exceeds unity for
$N \geq 6$ for the entire dynamic range plotted.  Furthermore,
Fig.~\ref{fig:dnr} indicates that even if the hierarchical ansatz is
correct (i.e., $S_N(R;z)$ is independent of $R$), the light-cone effect
should generate apparent scale-dependence, since the correction
behaves differently at different scales for a given redshift.  In
future surveys extending to $z>1$, Fig.~ \ref{fig:dnr} implies that
the required correction for the light-cone effect is appreciable,
ranging from up to unity for $S_3$ through factors of few for $S_6$ to
factors of hundred for $S_{10}$.

\section{Cosmological redshift-space distortion \label{sec:cosred}}

\subsection{Basic idea of cosmological redshift-space distortion  
\label{subsec:ideacrd}}

The approach described in \S \ref{sec:lightcone} is based on an implicit idea to
treat all objects in a survey catalogue simultaneously. If the number of
objects in the catalogue is sufficiently large, one can divide the
objects in many redshift bins. Then the light-cone effect discussed
above is less important, as long as one treats each individual bin
separately. In this case, however, another interesting effect due to the
geometry of the universe emerges. This originates from the fact that the
(observable) redshift-space separation $\bf s$ is mapped to the
(unobservable) comoving separation $\bf x$ of objects at $z$ differently,
depending on whether the separation is parallel or perpendicular to the
line-of-sight direction of an observer at $z=0$. Due to this effect, a
sphere located at $z$ becomes elongated along the line-of-sight in
general.\cite{AP} In this section, we describe the anisotropy in the
two-point correlation function of high-redshift objects induced by this
effect, which we call the cosmological redshift-space
distortion,\cite{MS96,BPH} in order to distinguish the conventional
redshift-space distortion due to the peculiar velocity
field.\cite{DP83}\tocite{Hamilton97}

%%%%%%%%%%%%%%%%%%%%%%%%%%%%%%%%%%%%%%%%%%%%%%%%%%%%%%%%%%%%%%%%%%%%%
\begin{figure}[bh]
\begin{center}
 \leavevmode\epsfysize=6cm \epsfbox{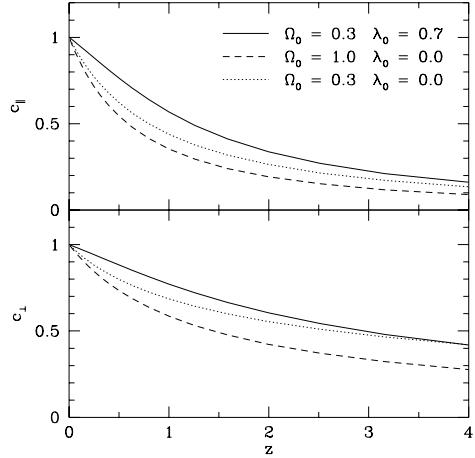}
\end{center}
\caption{Behavior of $c_\spara(z)$ and $c_\sperp (z)$
  for ($\Omega_0$, $\lambda_0$) = (0.3,0.7),
  (1.0,0.0), and (0.3,0.0) corresponding to the solid, dashed and
  dotted lines, respectively.
\label{fig:cz}}
\end{figure}
%%%%%%%%%%%%%%%%%%%%%%%%%%%%%%%%%%%%%%%%%%%%%%%%%%%%%%%%%%%%%%%%%%%%%

Throughout this section, we assume a standard Robertson -- Walker metric
of the form
%%%%%%%%%%%%%%%%%%%%%%%%%%%%%%%%%%%%%%%%%%%%%%%%%%%%%%%%%%%%%%%%%%%
\begin{equation}
ds^2 = -dt^2 + a(t)^2 
\{ d\chi^2 + S(\chi)^2 [d\theta^2 + \sin^2\theta d\phi^2 ] \} .
\end{equation}
%%%%%%%%%%%%%%%%%%%%%%%%%%%%%%%%%%%%%%%%%%%%%%%%%%%%%%%%%%%%%%%%%%%
We adopt a normalization for which the present scale factor $a_0$ is
unity. Then the spatial curvature $K$ is related to the other parameters
as
%%%%%%%%%%%%%%%%%%%%%%%%%%%%%%%%%%%%%%%%%%%%%%%%%%%%%%%%%%%%%%%%%%%
\begin{equation}
K = H_0^2 (\Omega_0 + \lambda_0 -1) ,
\end{equation}
%%%%%%%%%%%%%%%%%%%%%%%%%%%%%%%%%%%%%%%%%%%%%%%%%%%%%%%%%%%%%%%%%%%
and $S(\chi)$ is determined by the sign of $K$ according to 
%%%%%%%%%%%%%%%%%%%%%%%%%%%%%%%%%%%%%%%%%%%%%%%%%%%%%%%%%%%%%%%%%%%
\begin{equation}
 S(\chi) = 
  \left\{ 
      \begin{array}{ll}
         \sin{(\sqrt{K}\chi)}/\sqrt{K} & \mbox{$(K>0)$} \\
         \chi & \mbox{$(K=0)$} \\
         \sinh{(\sqrt{-K}\chi)}/\sqrt{-K} & \mbox{$(K<0)$} 
      \end{array}
   \right. .
 \\ \nonumber
\end{equation}
%%%%%%%%%%%%%%%%%%%%%%%%%%%%%%%%%%%%%%%%%%%%%%%%%%%%%%%%%%%%%%%%%%%
The radial distance $\chi(z)$ is given by
%%%%%%%%%%%%%%%%%%%%%%%%%%%%%%%%%%%%%%%%%%%%%%%%%%%%%%%%%%%%%%%%%%%
\begin{equation}
\chi(z) = \int_t^{t_0} {dt \over a(t)}  
 =  \int_0^z {d z \over H(z)} .
\end{equation}
%%%%%%%%%%%%%%%%%%%%%%%%%%%%%%%%%%%%%%%%%%%%%%%%%%%%%%%%%%%%%%%%%%%

%%%%%%%%%%%%%%%%%%%%%%%%%%%%%%%%%%%%%%%%%%%%%%%%%%%%%%%%%%%%%%%%%%%%%
\begin{figure}[hb]
\begin{center}
   \leavevmode\epsfysize=8cm \epsfbox{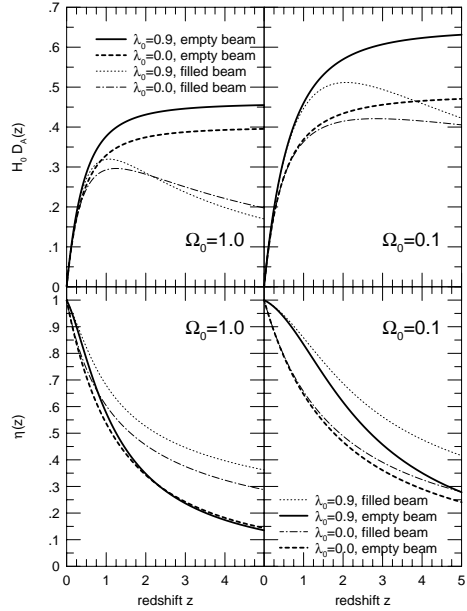}
\end{center}
\caption{ Effect of inhomogeneity on the angular diameter distance
  $D_\A(z)$ (upper panels) and the correction factor
  $\eta(z)=c_\spara(z)/c_\sperp (z)$ (lower panels) for $\lambda_0=0$ and
  $0.9$ models in $\Omega_0=1$ (left panels) and $\Omega_0=0.1$ (right
  panels) universes. Thick lines indicate the results for the empty beam
  ($\alpha=0)$, while thin lines for the filled beam ($\alpha=1)$.
  \label{fig:daeta} }
\end{figure}
%%%%%%%%%%%%%%%%%%%%%%%%%%%%%%%%%%%%%%%%%%%%%%%%%%%%%%%%%%%%%%%%%%%%%

Consider a pair of objects located at redshifts $z_1$ and
$z_2$.  If both the redshift difference $\delta z \equiv z_1-z_2$ and
the angular separation of the pair $\delta\theta$ are much less than
unity, the comoving separations of the pair parallel and perpendicular
to the line-of-sight direction, $x_{\spara}$ and $x_\sperp$, are given
by
%%%%%%%%%%%%%%%%%%%%%%%%%%%%%%%%%%%%%%%%%%%%%%%%%%%%%%%%%%%%%%%%%%%
\begin{equation}
\label{eq:xdef2}
x_{\spara} (z) = {d\chi(z)\over dz}  \delta z =
 \frac{c_{\spara} \delta z}{H_0}, \quad
x_\sperp (z)  = S(\chi(z)) \delta \theta =
\frac{c_\sperp z \delta \theta}{H_0},
\end{equation}
%%%%%%%%%%%%%%%%%%%%%%%%%%%%%%%%%%%%%%%%%%%%%%%%%%%%%%%%%%%%%%%%%%%
where $c_\spara (z) = H_0/H(z)$, $c_\sperp (z) = H_0 S(\chi(z))/z$
and $z \equiv (z_1+z_2)/2$. Thus their ratio becomes
%%%%%%%%%%%%%%%%%%%%%%%%%%%%%%%%%%%%%%%%%%%%%%%%%%%%%%%%%%%%%%%%%%%
\begin{eqnarray} 
\label{eq:ratio}
{x_\spara (z) \over x_\sperp (z) } 
= \frac{c_\spara(z)}{c_\sperp (z)}{\delta z \over z \delta \theta}
\equiv \eta(z) {\delta z \over z \delta \theta} .
\end{eqnarray}
%%%%%%%%%%%%%%%%%%%%%%%%%%%%%%%%%%%%%%%%%%%%%%%%%%%%%%%%%%%%%%%%%%%
Since $\delta z/(z \delta \theta)$ is the ratio of the parallel and
perpendicular separations to the line-of-sight direction, $\eta(z)$
represents the distortion in the redshift space coordinates induced by
the geometry of the universe. Typical behavior of $c_\spara (z)$,
$c_\sperp (z)$ and $\eta(z)$ is plotted in Fig.~\ref{fig:cz} and the lower
panels of Fig.~\ref{fig:daeta}. This is purely a general relativistic
effect which was first proposed as a potential test of the
cosmological constant.\cite{AP}

Actually it took a couple of decades to find realistic phenomena to
which the test could be applied observationally.  Recently two independent
groups\cite{MS96,BPH} proposed to use the anisotropy in the clustering
pattern of quasars and galaxies at high-redshifts as a probe
of $\eta(z)$. The next subsections describe the methodology in linear
theory, and then examine the feasibility in the nonlinear regime using
$N$-body simulations.\cite{Magira}

\subsection{Linear theory of cosmological redshift-space distortion
 \label{subsec:linearcrd}}

We choose a fiducial point at redshift $z$ as an origin, and set up locally
Euclidean coordinates with respect to this point. Let us adopt the
distant (plane-parallel) observer approximation and choose the
line-of-sight direction as the third axis.  Then, if an object is located
at $\bfx=(x_1,x_2,x_3)$ in the real (comoving) space, the corresponding
redshift-space coordinates, $\bfs=(s_1,s_2,s_3)$, observed at $z=0$ are
written as
%%%%%%%%%%%%%%%%%%%%%%%%%%%%%%%%%%%%%%%%%%%%%%%%%%%%%%%%%%%%%%%%%%%
\begin{subeqnarray}
  s_1 &=& \frac{x_1}{c_\sperp (z)}, \\
  s_2 &=& \frac{x_2}{c_\sperp (z)}, \\
  s_3 &=& \frac{z_{\rm obs} - z}{H_0} \simeq
   {1 \over c_\spara(z)} \left[x_3 + 
          \frac{1 + z}{H(z)}v_{\spara} \right] .
\label{eq:x2s}
\end{subeqnarray}
%%%%%%%%%%%%%%%%%%%%%%%%%%%%%%%%%%%%%%%%%%%%%%%%%%%%%%%%%%%%%%%%%%%
In the last expression, $v_{\spara}$ is the recession velocity of the
object relative to the observer, and $z \ll H(z)x_3$ is assumed.
Computing the Jacobian of the above transformation in linear theory, one
can relate the density contrasts of the objects in real and redshift
spaces as
%%%%%%%%%%%%%%%%%%%%%%%%%%%%%%%%%%%%%%%%%%%%%%%%%%%%%%%%%%%%%%%%%%%
\begin{equation}
   \delta^{(s)} (\bfs(\bfx)) = 
   \delta^{(r)} (\bfx) -  \frac{1 + z}{H(z)} \partial_3 v_\spara .
   \label{eq:dsdr}
\end{equation}
%%%%%%%%%%%%%%%%%%%%%%%%%%%%%%%%%%%%%%%%%%%%%%%%%%%%%%%%%%%%%%%%%%%
The peculiar velocity in linear theory is written in terms of the {\it
  mass} density contrast $\delta_\mass$ as\cite{Peebles93}
%%%%%%%%%%%%%%%%%%%%%%%%%%%%%%%%%%%%%%%%%%%%%%%%%%%%%%%%%%%%%%%%%%%
\begin{equation}
  v_\spara (\bfx) = - \frac{H(z)}{1 + z} f(z) \partial_3
  \triangle^{-1} \delta_\mass(\bfx),
\end{equation}
%%%%%%%%%%%%%%%%%%%%%%%%%%%%%%%%%%%%%%%%%%%%%%%%%%%%%%%%%%%%%%%%%%%
where $\triangle^{-1}$ is the inverse Laplacian, and
%%%%%%%%%%%%%%%%%%%%%%%%%%%%%%%%%%%%%%%%%%%%%%%%%%%%%%%%%%%%%%%%%%%
\begin{subeqnarray} 
\label{eq:fz}
   f(z) &\equiv& \frac{d\ln D(z)}{d\ln a} \simeq
   \Omega(z)^{0.6} + {\lambda(z) \over 70} 
\left(1+ {\Omega(z) \over 2}\right), \\
   \Omega(z) &=& \left[{H_0 \over H(z)}\right]^2 \, (1+z)^3 \Omega_0 ,\\
   \lambda(z) &=& \left[{H_0 \over H(z)}\right]^2 \, \lambda_0 .
\end{subeqnarray}
%%%%%%%%%%%%%%%%%%%%%%%%%%%%%%%%%%%%%%%%%%%%%%%%%%%%%%%%%%%%%%%%%%%

%%%%%%%%%%%%%%%%%%%%%%%%%%%%%%%%%%%%%%%%%%%%%%%%%%%%%%%%%%%%%%%%%%%%
\begin{figure}[h]
\begin{center}
 \leavevmode\epsfysize=7.0cm \epsfbox{xicontcl_g02s1.cps}
\end{center}
\caption{Contours of $\xi^{(s)}(s_\sperp, s_{\spara})$ for the fixed
  shape parameter $\Gamma$ ($=0.2$) and $\sigma_8=1.0$ models at $z=0.2$
  ({\it upper panels}) and $z=2.0$ ({\it lower panels}); $(\Omega_0,
  \lambda_0)= (1.0, 0.0)$, $(1.0, 0.9)$, $(0.1, 0.0)$ and $(0.1, 0.9)$
  from left to right. Solid and dashed lines indicate the
  positive and negative $\xi^{(s)}$, respectively.  Contour spacings are
  $\Delta {\rm log}_{10} |\xi| = 0.25$. \label{fig:xicontclg02}}
\vspace*{0.5cm}
\begin{center}
 \leavevmode\epsfysize=7.0cm \epsfbox{xicontcl_cdmg02.cps}
\end{center}
\caption{Contours of $\xi^{(s)}(s_\sperp, s_{\spara})$ at $z=3$.  Upper
  and lower panels correspond to the COBE-normalized CDM models (the
  shape parameter $\Gamma$ is given by
  Eq.~(\protect\ref{eq:gamma}\protect)), the fixed $\Gamma$ ($=0.2$)
  models with $\sigma_8=1.0$. Contour spacings and line types are the
  same as in Fig.~\protect\ref{fig:xicontclg02}\protect.
\label{fig:xicontclcdmg02}}
\end{figure}
%%%%%%%%%%%%%%%%%%%%%%%%%%%%%%%%%%%%%%%%%%%%%%%%%%%%%%%%%%%%%%%%%%%%%%

In order to close the equations, one has to relate the density contrast
of objects in real space $\delta^{(r)}$ to that of {\it mass},
$\delta_\mass$, by specifying the model of bias. As in \S
\ref{subsec:predictxi}, we adopt a linear bias:
%%%%%%%%%%%%%%%%%%%%%%%%%%%%%%%%%%%%%%%%%%%%%%%%%%%%%%%%%%%%%%%%%%%
\begin{equation} 
\label{eq:linearbias}
\delta^{(r)}(\bfx;z) = b(z) \delta_\mass(\bfx;z) .
\end{equation}
%%%%%%%%%%%%%%%%%%%%%%%%%%%%%%%%%%%%%%%%%%%%%%%%%%%%%%%%%%%%%%%%%%%
Substituting the above equations into Eq. (\ref{eq:dsdr}), we
obtain
%%%%%%%%%%%%%%%%%%%%%%%%%%%%%%%%%%%%%%%%%%%%%%%%%%%%%%%%%%%%%%%%%%%%%
\begin{equation}
\label{eq:kaiserz}
   \delta^{(s)}(\bfs(\bfx)) =
   \int \frac{d^3k}{(2\pi)^3}
   \left[1 + \beta(z) \frac{k_3^{\;2}}{k^2} \right]
   e^{i \bfk \cdot \bfx}
   {\widetilde \delta}^{(r)}(\bfk),
\end{equation}
%%%%%%%%%%%%%%%%%%%%%%%%%%%%%%%%%%%%%%%%%%%%%%%%%%%%%%%%%%%%%%%%%%%%%
where $\beta(z) = f(z)/b(z)$, and ${\widetilde \delta}^{(r)}$ is the
Fourier transform of $\delta^{(r)}$. Repeating the method of Hamilton
(1992),\cite{Hamilton92} we obtain an explicit formula for the
redshift-space two-point correlation function which is valid at
arbitrary $z$ in linear theory:
%%%%%%%%%%%%%%%%%%%%%%%%%%%%%%%%%%%%%%%%%%%%%%%%%%%%%%%%%%%%%%%%%%%
\begin{subeqnarray} 
\label{eq:xis}
   \xi^{(s)}(s_\sperp, s_\spara) &=& 
   \left(1+{2\over 3}\beta(z) +{1\over5}[\beta(z)]^2\right) \xi_0(x)
   P_0(\mu) \cr 
   &-& \left({4\over 3}\beta(z)
   +{4\over7}[\beta(z)]^2\right)\xi_2(x) P_2(\mu) \cr
   &+&\frac{8}{35}[\beta(z)]^2 \xi_4(x) P_4(\mu) , \\
\xi_{2l}(x) &=& {1 \over 2\pi^2} 
   \int_0^\infty dk k^2 j_{2l}(kx) P(k;z) , \\
  P(k;z) &=& [b(z)]^2\left[\frac{D(z)}{D(0)}\right]^2 P^\mass(k;z=0).
\end{subeqnarray}
%%%%%%%%%%%%%%%%%%%%%%%%%%%%%%%%%%%%%%%%%%%%%%%%%%%%%%%%%%%%%%%%%%%
In the above, $x \equiv \sqrt{{c_\spara}^2 {s_\spara}^2 + {c_\sperp}^2
  {s_\sperp}^2}$, $\mu\equiv c_{\spara} s_{\spara} /x$ ($s_\spara=s_3$
and $s_\sperp^2 = s_1^2+s_2^2$), the $P_n$ are the Legendre polynomials,
i.e., $P_0(\mu)=1$, $P_2(\mu)= (3\mu^2-1)/2$, and $P_4(\mu)=
(35\mu^4-30\mu^2+3)/8$. Figure~\ref{fig:xicontclg02} plots
$\xi^{(s)}(s_\sperp, s_{\spara})$ for $\Gamma=0.2$ and $\sigma_8=1.0$
models to illustrate the degree of distortion.
Figure~\ref{fig:xicontclcdmg02} shows the difference between CDM models
with $\Gamma$ of Eq. (\ref{eq:gamma}) and fixed $\Gamma (=0.2)$
models with the same $\Omega_0$ and $\lambda_0$.

\subsection{Implication for galaxy redshift surveys
\label{subsec:betacrd}}

The cosmological distortion effect becomes important also even for
shallower galaxy redshift surveys.\cite{GW} One may formally expand
$\xi^{(s)}(s_\sperp, s_{\spara})$ in terms of the {\it observables}, $s
\equiv \sqrt{s_{\spara}^2 + s_\sperp^2}$ and $\mu_s\equiv s_{\spara}/s$,
instead of the {\it unobservable} variables ($x(z)$, $\mu_x(z)$):
%%%%%%%%%%%%%%%%%%%%%%%%%%%%%%%%%%%%%%%%%%%%%%%%%%%%%%%%%%%%%%%%
\begin{equation}
\xi^{(s)}(s_\sperp, s_{\spara};z) 
= \sum_{l=0}^2 \xi_{2l}(x;z)
P_{2l}(\mu_x)
= \sum_{l=0}^\infty \zeta^{(s)}_{2l}(s;z) P_{2l}(\mu_s) .
\end{equation}
%%%%%%%%%%%%%%%%%%%%%%%%%%%%%%%%%%%%%%%%%%%%%%%%%%%%%%%%%%%%%%%%%
Since we are interested in surveys for which $z\ll 1$, we can further expand
the above summation up to first order in $z$, and then obtain
%%%%%%%%%%%%%%%%%%%%%%%%%%%%%%%%%%%%%%%%%%%%%%%%%%%%%%%%%%%%
\begin{equation}
\xi^{(s)}(s_\sperp, s_{\spara};z) 
  \approx \sum_{l=0}^3 \zeta^{(s)}_{2l}(s;z) P_{2l}(\mu_s) .
\end{equation}
%%%%%%%%%%%%%%%%%%%%%%%%%%%%%%%%%%%%%%%%%%%%%%%%%%%%%%%%%%%%
The explicit expressions for $\zeta^{(s)}_{2l}(s;z)$ of the form:
%%%%%%%%%%%%%%%%%%%%%%%%%%%%%%%%%%%%%%%%%%%%%%%%%%%%%%%%%%%%%%%%%%%
\begin{equation} 
\label{eq:xi2zeta}
\zeta^{(s)}_{2l}(x;z) = \xi^{(s)}_0(x;0) + \delta_{2l} z 
\end{equation}
%%%%%%%%%%%%%%%%%%%%%%%%%%%%%%%%%%%%%%%%%%%%%%%%%%%%%%%%%%%%%%%%%%%
for $l=0$, 1, 2 and 3 are found in Ref.~\citen{NMS}.

One possible application of those perturbative formulae is to estimate
a systematic error for the $\beta$-parameter, $\beta_0$, due to the
neglect of the cosmological redshift-space distortion.  Hamilton
(1992) proposed to estimate $\beta_0$ from the moments of the observed
two-point correlation functions of galaxies on the basis of the
relation\cite{Hamilton92} 
%%%%%%%%%%%%%%%%%%%%%%%%%%%%%%%%%%%%%%%%%%%%%%%%%%%%%%%%%%%%%%%%
\begin{equation}
\frac{1+2\beta_0/3 +\beta_0^2/5}{4\beta_0/3+4\beta_0^2/7}
= \frac{\xi^{(s)}_0(r;0)}{\xi^{(s)}_2(r;0)}
- 3 \int_0^r \frac{\xi^{(s)}_0(x;0)}{\xi^{(s)}_2(r;0)}
      \left({x\over r}\right)^3 {dx \over x} .
\label{eq:H92beta0}
\end{equation}
%%%%%%%%%%%%%%%%%%%%%%%%%%%%%%%%%%%%%%%%%%%%%%%%%%%%%%%%%%%%
If one takes account of the cosmological redshift-space distortion at
$z$, $\xi^{(s)}_0(x;0)$ and $\xi^{(s)}_2(x;0)$ on the right-hand side
of the above equation should be replaced by $\zeta^{(s)}_0(x;z)$ and
$\zeta^{(s)}_2(x;z)$, respectively. Then substituting the perturbation
expansion (\ref{eq:xi2zeta}) into Eq. (\ref{eq:H92beta0}), one
can compute the systematic error for $\beta_0$ defined through
%%%%%%%%%%%%%%%%%%%%%%%%%%%%%%%%%%%%%%%%%%%%%%%%%%%%%%%%%%%%%%%%%%%
\begin{equation} 
\beta(z) = \beta_0 + \epsilon z .
\end{equation} 
%%%%%%%%%%%%%%%%%%%%%%%%%%%%%%%%%%%%%%%%%%%%%%%%%%%%%%%%%%%%%%%%%%%
The result\cite{NMS} consists of the two terms corresponding to the
evolution of the $\beta$-parameter and the geometrical effect:
%%%%%%%%%%%%%%%%%%%%%%%%%%%%%%%%%%%%%%%%%%%%%%%%%%%%%%%%%%%%
\begin{eqnarray}
&&
\hspace*{-1.2cm}
\epsilon
= \left.{d\,\beta(z) \over d z}\right|_{z=0} 
+ {1+q_0 \over 1+ 6\beta_0/7 + 3\beta_0^2/35}
\times
\left[{\beta_0 \over 7} 
\left(1-{\beta_0\over 5}-{11\beta_0^2 \over 15}-{\beta_0^3\over 7}
\right) 
\right. 
\nonumber \\
&&
\hspace*{2cm}
\left. 
+ \frac{1/4 + 3\beta_0/7 + 17\beta_0^2/70 + \beta_0^3/21 + \beta_0^4/196}
{{3 \over r^3} \int_0^r [\xi(x;0)-\xi(r;0)]x^2dx} \,
{\partial\xi(r;0) \over \partial\ln r}
\right] ,
\end{eqnarray}
%%%%%%%%%%%%%%%%%%%%%%%%%%%%%%%%%%%%%%%%%%%%%%%%%%%%%%%%%%%%
where $q_0$ is the deceleration parameter ($ = \Omega_0/2 -
\lambda_0$). For magnitude-limited samples, the above expression
should be averaged over the sample with an appropriate weight
according to the selection function.  If the magnitude limit of the
survey is 18.5 (in B-band) as in the SDSS galaxy redshift sample, the
systematic error ranges between $-20\%$ and $10\%$, depending on the
cosmological parameters. Although such systematic errors are smaller
than the statistical errors in the current surveys, they will
definitely dominate the expected statistical error for future surveys.

\subsection{Effects of the cosmological distance and evolution of bias
\label{subsec:alphacrd}}

Here we discuss two potentially important effects, which were ignored in
\S \ref{subsec:ideacrd} to \S \ref{subsec:betacrd}, on the cosmological
redshift-space distortion: the effect of inhomogeneities in the light
propagation and the evolution of bias.

The angular diameter distance $D_\A$, which plays a key role in the
geometrical test at high $z$, depends sensitively on the inhomogeneous
matter distribution as well as $\lambda_0$ and the density parameter
$\Omega_0$.  A reasonably realistic approximation to the light
propagation in an inhomogeneous universe is given by the Dyer -- Roeder
distance.\cite{DR73} It assumes that the fraction $\alpha$ of the total
matter in the universe is distributed smoothly and the rest is in 
clumps. If an observed beam of light propagates far from any clump,
then the angular diameter distance $D_\A(z;\alpha,\Omega_0,\lambda_0)$
satisfies
%%%%%%%%%%%%%%%%%%%%%%%%%%%%%%%%%%%%%%%%%%%%%%%%%%%%%%%%%%%%%%%%%%%%%%%%
\begin{equation}
\label{eq:da}
{d^2 D_\A \over dz^2} 
+ \left[ {2 \over 1+z} + {1\over H(z)} {d H(z) \over dz}\right]
{d D_\A \over dz} 
+ {3 \over 2} {\alpha H_0^2 \Omega_0 (1+z) \over H(z)^2} D_\A = 
0 ,
\end{equation}
%%%%%%%%%%%%%%%%%%%%%%%%%%%%%%%%%%%%%%%%%%%%%%%%%%%%%%%%%%%%%%%%%%%%%%%%
with $D_\A(z=0)=0$ and $d D_\A/dz (z=0)= 1/H_0$.  The preceding
discussion on the cosmological redshift-space distortion adopted a
{\it standard} distance, which corresponds to the extreme case of
$\alpha=1$. As shown in Fig.~\ref{fig:daeta}, the effect of
inhomogeneity represented by the parameter $\alpha$ in the above
approximation, however, could be large for high $z$ if $\alpha$ is
significantly different from unity.  Another uncertainty will come
from the possible time-dependence of the bias parameter $b(z)$. As we
emphasized in the discussion of the light-cone effect, we do not yet have
any reliable theoretical model for bias. Thus we adopt the linear
bias model (\ref{eq:fryb}) so as to highlight the effect of the
evolution of bias in the present context.

%%%%%%%%%%%%%%%%%%%%%%%%%%%%%%%%%%%%%%%%%%%%%%%%%%%%%%%%%%%%%%%%%%%%
\begin{figure}[tbh]
\begin{center}
   \leavevmode\epsfysize=6cm \epsfbox{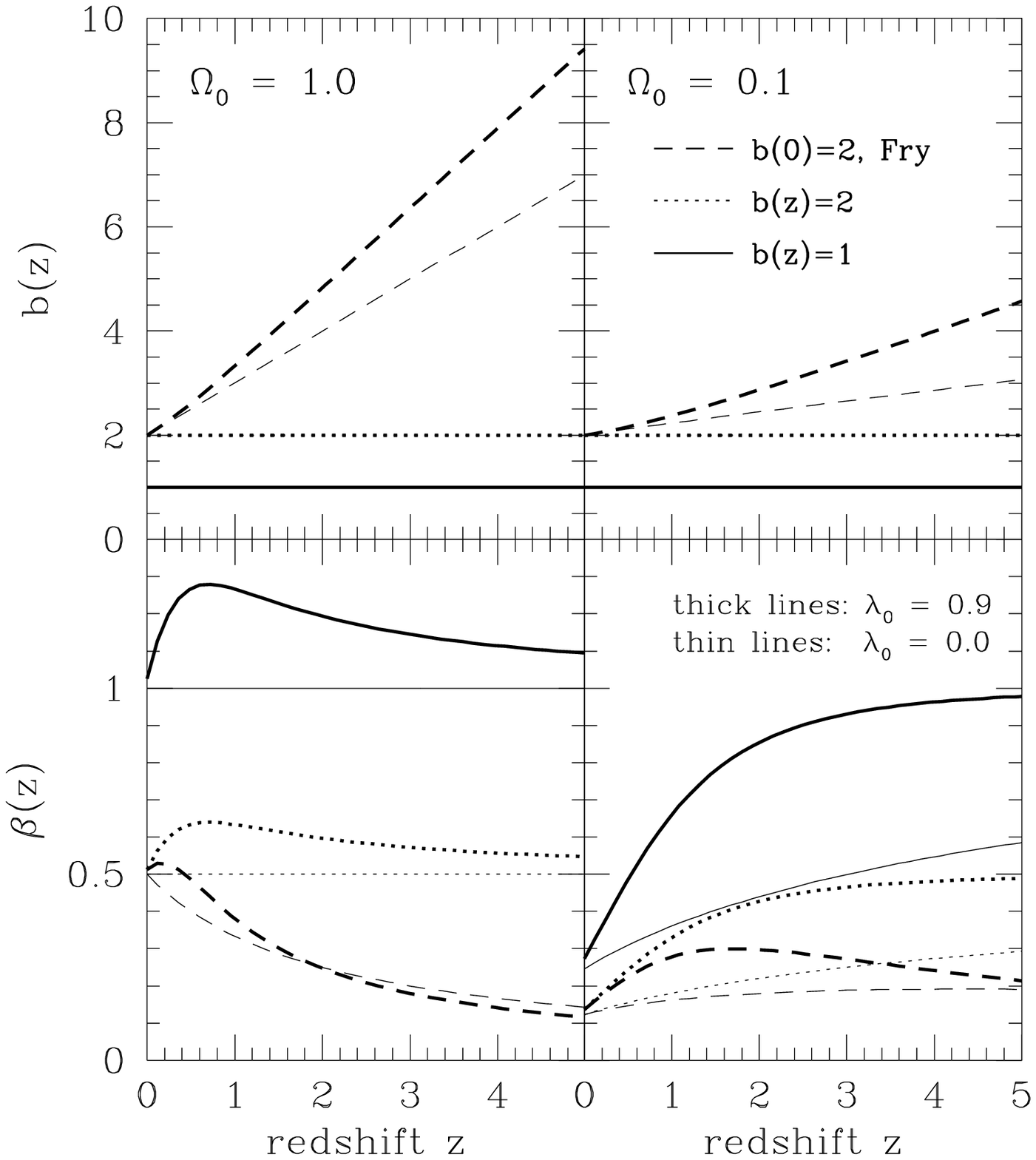}
\end{center}
\caption{
  Evolution of bias and $\beta(z)$. Dashed and solid lines correspond to
  constant biasing parameter $b(z)=2$ and $1$, respectively, while
  dotted lines correspond to the bias model
  (Eq.~(\protect\ref{eq:fryb}\protect)) with $b=2$ at $z=0$. Thick and
  thin lines correspond to $\lambda_0=0.9$ and $0$, respectively.
\label{fig:bbeta}
}\end{figure}
%%%%%%%%%%%%%%%%%%%%%%%%%%%%%%%%%%%%%%%%%%%%%%%%%%%%%%%%%%%%%%%%%%%%%%

Figure~\ref{fig:daeta} may seem to indicate that inhomogeneity makes
even a larger difference than that of $\lambda_0$, especially for $z\gg
1$.  In reality, however, the situation is not so bad; since the
expectation value of $\alpha$ is determined by the effective volume of
the beam of the light bundles, it depends on the depth $z$ and the
angular separation $\delta \theta$ (of the quasar pair in the present
example). For larger $z$ and larger $\delta \theta$, $ \alpha (z,
\delta\theta)$ should approach unity in any case, and the result based
on the standard distance as in \S \ref{subsec:ideacrd} to \S
\ref{subsec:betacrd} would be more appropriate and closer to the
truth.  Since we do not have any justifiable model for $ \alpha (z,
\delta\theta)$, we will consider two extreme cases, $\alpha (z,
\delta\theta) = 1$ (filled beam) and $0$ (empty beam). Our main
purpose here is to highlight the importance of the effect, even though
more realistically $\alpha (z, \delta\theta)$ is somewhere between
the two extreme cases; it is shown that $\alpha (z, \delta\theta)=1$
is a good approximation for $z\gg1$ and $\delta\theta
\gg1$.\cite{tomita}

It is quite reassuring that even in these extreme cases the
inhomogeneity effect is much weaker than that of $\lambda_0$ up to
$z\simlt 2$ in {\it low density} universes, as the right panels in
Fig.~\ref{fig:daeta} illustrate clearly. Since a relatively low value
of $\Omega_0$ around $0.1- 0.3$ is favored
observationally,\cite{Peebles93} the optimal redshift to determine
$\lambda_0$ in low $\Omega_0$ universes is $z=1-2$.

%%%%%%%%%%%%%%%%%%%%%%%%%%%%%%%%%%%%%%%%%%%%%%%%%%%%%%%%%%%%%%%%%%%%%
\begin{figure}[tbh]
  \parbox{\halftext}{
\begin{center}
   \leavevmode\epsfysize=6cm \epsfbox{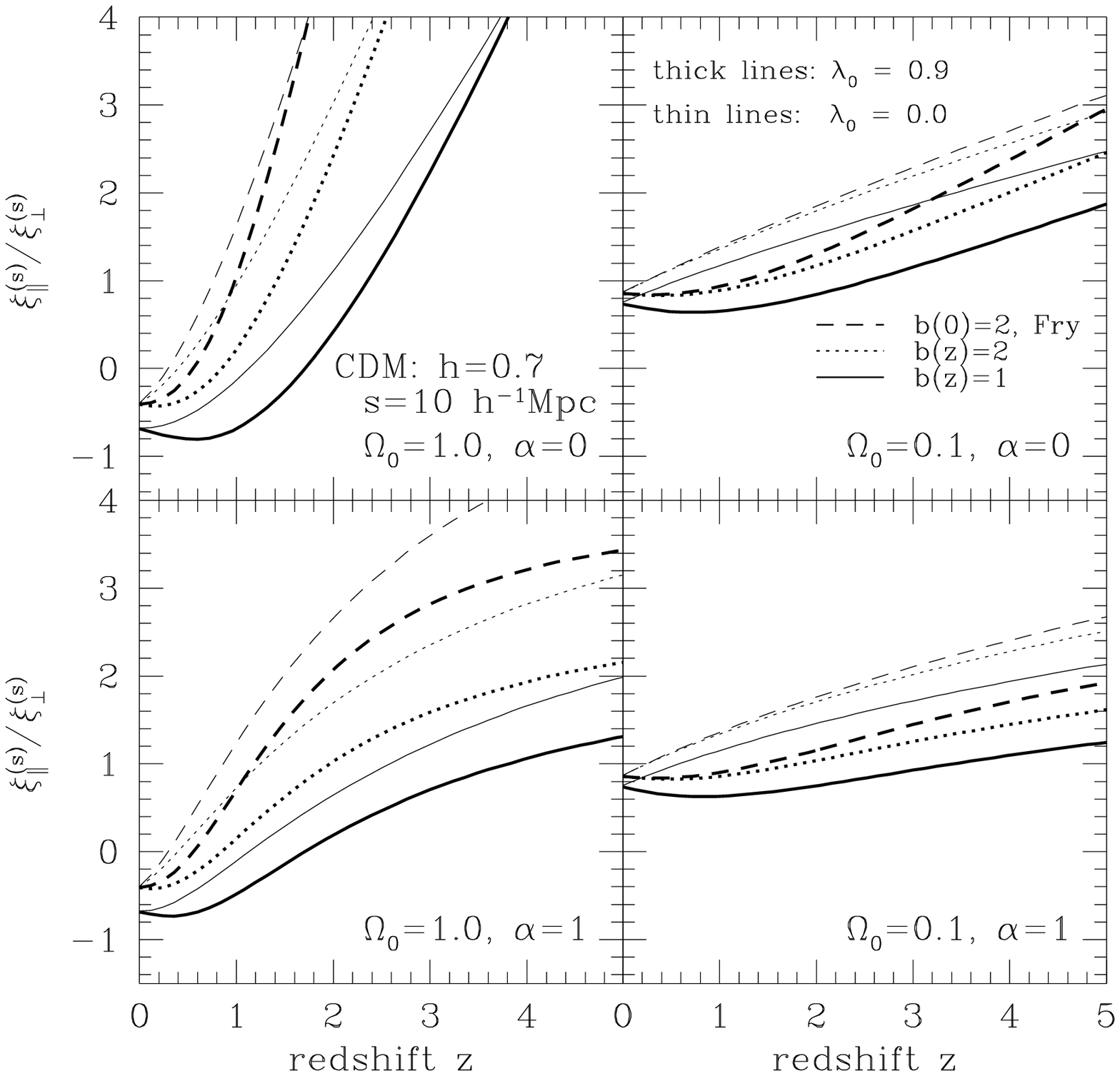}
\end{center}
\caption{The anisotropy parameter
  $\xi^{(s)}_{\spara}(s)/\xi^{(s)}_\sperp(s)$ as a function of $z$ at
  $s=10h^{-1}$Mpc in cold dark matter universes with
  $H_0=70$km/sec/Mpc.  The upper panels correspond to $\alpha=0$ (empty beams),
  while the lower panels correspond to $\alpha=1$ (filled beams).
\label{fig:anisoz}
}}
\hspace{0.4cm} 
\parbox{\halftext}{
\begin{center}
   \leavevmode\epsfysize=6cm \epsfbox{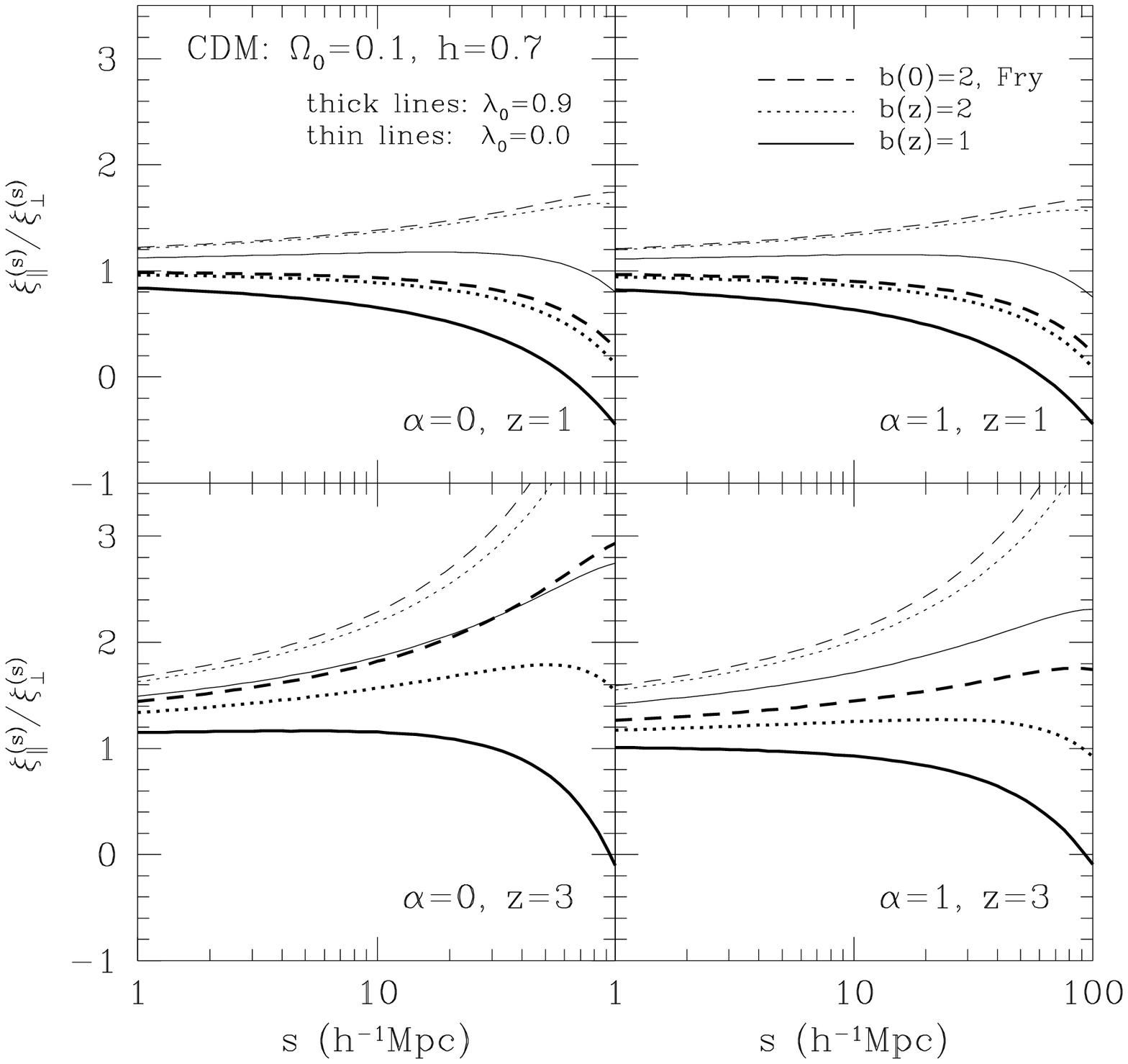}
\end{center}
\caption{The anisotropy parameter
  $\xi^{(s)}_{\spara}(s)/\xi^{(s)}_\sperp(s)$ as a function of $s$ at
  $z=1$ (upper panels) and at $z=3$ (lower panels) in cold dark matter
  universes with $H_0=70$km/sec/Mpc.  The left panels correspond to $\alpha=0$
  (empty beams), while the right panels correspond to $\alpha=1$ (filled beams).
\label{fig:anisos}
}}
\end{figure}
%%%%%%%%%%%%%%%%%%%%%%%%%%%%%%%%%%%%%%%%%%%%%%%%%%%%%%%%%%%%%%%%%%%%%

Figure~\ref{fig:bbeta} illustrated the evolution of bias
(Eq.~(\ref{eq:fryb}); upper panels) and of the resulting $\beta(z)$
parameter (lower panels). This implies that as long as Fry's model of
$b(z)$ is adopted, one can distinguish the value of $\lambda_0$
independently of the evolution of bias {\it only in low density
  ($\Omega_0 \ll 1$) models and at intermediate redshifts ($z\simlt
  2$)}. Together with the indication from Fig.~\ref{fig:daeta},
$z=1-2$ would be an optimal regime to probe $\lambda_0$ at least
in low-density universes.  Figure~\ref{fig:anisoz} illustrates the
extent to which this is feasible simply on the basis of the anisotropy
parameter,
%%%%%%%%%%%%%%%%%%%%%%%%%%%%%%%%%%%%%%%%%%%%%%%%%%%%%%%%%%%%%%%%%%%%%
\begin{equation}
{\xi^{(s)}_{\spara}(s) \over \xi^{(s)}_\sperp(s)}
\equiv {\xi^{(s)}(s_\spara=0,s;z) \over \xi^{(s)}(s,s_\sperp=0;z)} ,
\end{equation}
%%%%%%%%%%%%%%%%%%%%%%%%%%%%%%%%%%%%%%%%%%%%%%%%%%%%%%%%%%%%%%%%%%%%%
adopting the power spectrum of the CDM models; in $\Omega_0=1$ models
the value of $\alpha$ completely changes the $z$-dependence of the
anisotropy parameter while $\Omega_0=0.1$ models are fairly insensitive
to it. In addition, $\xi^{(s)}_{\spara}(s)/\xi^{(s)}_\sperp(s)$ for
$z\simlt 2$ in $\Omega_0=0.1$ models is basically determined by the
biasing parameter at $z=0$ and less affected by the evolution of $b(z)$.
Figure~\ref{fig:anisos} shows the scale-dependence of the anisotropy
parameter in $\Omega_0=0.1$ and $h=0.7$ CDM models.  This clearly
indicates that one can distinguish the different $\lambda_0$ and bias
models by analyzing the anisotropy of the correlation function at $z=1$
almost independently of $\alpha$.

\subsection{Testing the redshift-space distortion with
 N-body simulations \label{subsec:nbodycrd}}

As illustrated in Figs.~\ref {fig:daeta} to \ref{fig:xicontclcdmg02},
the two-point correlation functions become elongated along the
line-of-sight due to the cosmological redshift-space distortion in
linear theory.  In reality, the {\it finger of god} due to the
non-linear peculiar velocity affects the distortion pattern in the
same direction. Therefore the proper modeling of the nonlinear effects
is essential to estimate the cosmological parameters from the observed
distortion. Also, the available number of observed objects would limit
the statistical significance of the analysis. In order to examine
these realistic effects in applying the redshift-space distortion as a
cosmological test, we use a series of high-resolution N-body
simulations.\cite{JS98,Jing,Magira} The simulations assume the three
representative cosmological models summarized in
Table~\ref{tab:modelparam}.  Each model has three realizations with
different random seeds in generating the initial condition, and
employs $N_P=256^3$ dark matter particles in the simulation volume of
($300h^{-1}$Mpc)$^3$ (comoving).  Figure~\ref{fig:xicontcl_nbodydmgp}
displays the results of $\xi^{(s)}(s_\sperp,s_{\spara})$ at $z=2.2$
for these models. The upper panels plot the predictions in linear theory,
the middle panels are computed from randomly sampled $N=5\times10^5$
particles, and the lower panels from the $N=2\times10^4$ most massive
halos (groups of particles) identified,\cite{JS98,Jing} so as to take
into account the effects of the finite sampling and the biasing to
some degree.

%%%%%%%%%%%%%%%%%%%%%%%%%%%%%%%%%%%%%%%%%%%%%%%%%%%%%%%%%%%%%%%%%%%%%%%%%%
\begin{table}[h]
\caption{Simulation model parameters. \label{tab:modelparam}}
\begin{center}
\begin{tabular}{ccccccccc}
\hline
Model & $\Omega_0$ &  $\lambda_0$  
&  $\Gamma$ &   $\sigma_8$  & N & realizations \\ 
\hline
SCDM & 1.0  & 0.0 & 0.5  & 0.6 & $256^3$ & 3  \\
OCDM & 0.3  & 0.0 & 0.25 & 1.0 & $256^3$ & 3 \\
LCDM & 0.3  & 0.7 & 0.21 & 1.0 & $256^3$ & 3 \\
\hline
\end{tabular}
\end{center}
\end{table}
%%%%%%%%%%%%%%%%%%%%%%%%%%%%%%%%%%%%%%%%%%%%%%%%%%%%%%%%%%%%%%%%%%%%%%%%%%

Figure~\ref{fig:omegalambda_nbody} plots the reduced $\chi^2$ contours
from $\xi^{(s)}(s_\sperp, s_{\spara})$ of simulations.  Since our
theoretical predictions do not include nonlinear effects at this
point, we exclude the regions with $s_{\spara}/s_\sperp>2$, which are
likely to be seriously contaminated by nonlinear peculiar velocities.
While Fig.~\ref{fig:omegalambda_nbody} demonstrates that the current
methodology works in principle, the expected S/N is fairly low.  This
is largely because we adjusted the sampling rate for the high-z QSOs.
The situation would be improved, though, if we apply the present
methodology to a statistical sample of Lyman-break galaxies, for
instance, whose number density is larger and their strong clustering
is already observed.\cite{Steidel}

%%%%%%%%%%%%%%%%%%%%%%%%%%%%%%%%%%%%%%%%%%%%%%%%%%%%%%%%%%%%%%%%%%%%%%%%
\begin{figure}[tbh]
\begin{center}
 \leavevmode\epsfysize=7.cm \epsfbox{xicontcl_nbodydmgp.cps}
\end{center}
\caption{ $\xi^{(s)}(s_\sperp, s_{\spara})$ at $z=2.2$ from linear
theory ({\it upper panels}), and N-body simulations with $5\times10^5$
randomly sampled particles ({\it middle panels}) and with
$N=2\times10^4$ most massive halos of particles ({\it lower
panels}). The contour lines represent $\xi^{(s)}=10$, $10^{0.5}$ and $1$
(in black), and $\xi^{(s)}=0.5$, $0.3$, $0.2$ and $0.1$ (in white).  The
white region around the upper-left corner in SCDM model indicates
$\xi^{(s)}<0$. \label{fig:xicontcl_nbodydmgp} }
\vspace*{0.5cm}
\begin{center}
  \leavevmode\epsfysize=5cm \epsfbox{omegalambda_nbody.cps}
\end{center}
\caption{ The resulting $\chi^2$-contours in the $\Omega_0$ - $\lambda_0$
  plane from the analysis of the data in
  Fig.~\protect\ref{fig:xicontcl_nbodydmgp}\protect. The theoretical
  predictions are not corrected for the nonlinear effects. The crosses
  indicate the true values adopted in the simulation models.
  \label{fig:omegalambda_nbody}}
\end{figure}
%%%%%%%%%%%%%%%%%%%%%%%%%%%%%%%%%%%%%%%%%%%%%%%%%%%%%%%%%%%%%%%%%%%%%%%%%

In order to examine the nonlinear effects in the cosmological
redshift-space distortion, we consider the power spectrum, rather than
the two-point correlation functions, in which the phenomenological
correction for the non-linear finger-of-god effect has already been discussed
in the literature.\cite{PD94,BPH} Specifically, we model the power
spectrum {\it before the cosmological redshift-space distortion} as
%%%%%%%%%%%%%%%%%%%%%%%%%%%%%%%%%%%%%%%%%%%%%%%%%%%%%%%%%%%%%%%%%%%%%%%%%%
\begin{equation}
  P^{(s)}(k,\mu)=P(k)[1+\beta\mu^2]^2 D[k\mu\sigma_P] ,
\end{equation}
%%%%%%%%%%%%%%%%%%%%%%%%%%%%%%%%%%%%%%%%%%%%%%%%%%%%%%%%%%%%%%%%%%%%%%%%%%
where $\mu$ is the direction cosine in $k$ space, and the second factor
on the right-hand-side comes from Eq. (\ref{eq:kaiserz}). The last
factor is a phenomenological correction for non-linear velocity effect. We
assume that the pair-wise velocity distribution in real space is
exponential with a constant pairwise peculiar velocity along the
line-of-sight, $\sigma_P$.  In this case the damping term in Fourier
space, $D[k\mu\sigma_P]$, is given by\cite{PD94}
%%%%%%%%%%%%%%%%%%%%%%%%%%%%%%%%%%%%%%%%%%%%%%%%%%%%%%%%%%%%%%%%%%%%%%%%%%
\begin{equation}
\label{eq:damping}
  D[k\mu\sigma_P]=\frac{1}{1+(k\mu\sigma_P)^2/2} .
\end{equation}
%%%%%%%%%%%%%%%%%%%%%%%%%%%%%%%%%%%%%%%%%%%%%%%%%%%%%%%%%%%%%%%%%%%%%%%%%%

Combining the geometrical effect, the power spectrum of objects at $z$
observed in redshift space is expressed as
%%%%%%%%%%%%%%%%%%%%%%%%%%%%%%%%%%%%%%%%%%%%%%%%%%%%%%%%%%%%%%%%%%%%%%%%%%
\begin{eqnarray}
\label{eq:pkcrd}
 P^{(CRD)}(k_{s\sperp},k_{s\spara};z)&=&
\frac{1}{c_\sperp(z)^2c_\spara(z)}P^{(s)}(k_\sperp,k_\spara;z) \cr
&=&\frac{1}{c_\sperp(z)^2c_\spara(z)}P(k;z)
   [1+\beta(z)\mu^2]^2D[k\mu\sigma_P(z)] ,\qquad
\end{eqnarray}
%%%%%%%%%%%%%%%%%%%%%%%%%%%%%%%%%%%%%%%%%%%%%%%%%%%%%%%%%%%%%%%%%%%%%%%%%%
where the relation of the comoving wave numbers in real space, $\bfk$, 
and in cosmological redshift space, $\bfk_s$, is expressed as
%%%%%%%%%%%%%%%%%%%%%%%%%%%%%%%%%%%%%%%%%%%%%%%%%%%%%%%%%%%%%%%%%%%%%%%%%%
\begin{eqnarray}
  k_\sperp = \frac{k_{s\sperp}}{c_\sperp(z)}, \qquad
  k_\spara=\frac{k_{s\spara}}{c_\spara(z)} , \qquad
  k = \sqrt{k_\sperp^2+k_\spara^2}, \qquad  \mu=\frac{k_\spara}{k},
\end{eqnarray}
%%%%%%%%%%%%%%%%%%%%%%%%%%%%%%%%%%%%%%%%%%%%%%%%%%%%%%%%%%%%%%%%%%%%%%%%%%
and $P(k;z)$ is a comoving real-space power spectrum at redshift $z$.

Clearly, the final expression for the redshift-space power spectrum
$P^{(CRD)}(k_{s\sperp},k_{s\spara};z)$ depends on a number of
parameters: $\Omega_0$, $\lambda_0$, $\sigma_8$, $b(z)$, $P(k;z)$, and
$\sigma_P(z)$. While none of these parameters has been determined precisely yet,
there exist some tight constraints on them which can greatly
reduce the number of the independent unknown parameters; provided that
one adopts the linear biasing, the shape of the linear density power
spectrum has already been determined fairly well by the APM galaxy
survey, for instance.\cite{BE} The upcoming redshift surveys of nearby
galaxies will improve this measurement significantly. Then, given
$\sigma_8$ and $b(z)$, $P(k;z)$ is already accurately determined.
Furthermore, with future large surveys which are pertinent to the
analysis here, it should be fairly easy to determine $b(z)$ for a given
cosmology. The pair-wise velocity dispersion $\sigma_P(R,z)$ at 
large separation can be determined by the other parameters through the
cosmic energy equation:\cite{MJB}
%%%%%%%%%%%%%%%%%%%%%%%%%%%%%%%%%%%%%%%%%%%%%%%%%%%%%%%%%%%%%%%%%%%
\begin{subeqnarray} \label{eq:spmjb}
  \hspace*{-1cm} 
\sigma_{\rm p,MJB} &\equiv&  \langle v_{12}^2(r\rightarrow\infty)\rangle 
= \frac23\langle v_1^2\rangle, \\
  \langle v_1^2\rangle &=& 
\frac{3\Omega(z)H^2(z)I_2(z)}{2(1+z)^2}
\left[1-\frac{1+z}{D^2(z)}\int_\frac{1}{1+z}^\infty 
   \frac{D^2(z')}{(1+z')^2}dz'\right] , \\
  I_2(z) &=& \int_0^{\infty}\frac{dk}{k}\frac{\Delta^2_\NL(k,z)}{k^2}
\end{subeqnarray}
%%%%%%%%%%%%%%%%%%%%%%%%%%%%%%%%%%%%%%%%%%%%%%%%%%%%%%%%%%%%%%%%%%%
%%%%%%%%%%%%%%%%%%%%%%%%%%%%%%%%%%%%%%%%%%%%%%%%%%%%%%%%%%%%%%%%%%%%%%%%%%
\begin{table}[htb]
\caption{Velocity dispersion $\sigma_P$ from theoretical 
  predictions and the present simulations (in units of km/sec).
\label{tab:sigmap}}
\begin{center}
\begin{tabular}{cllll}
  \hline Model & $\sigma_{P,{\rm MJB}}(z=0)$ & $\sigma_{P,{\rm
      sim}}(z=0)$ & $\sigma_{P,{\rm MJB}}(z=2.2)$ & $\sigma_{P,{\rm
      sim}}(z=2.2)$\\ \hline SCDM & 580 & $591 \pm 2$ & 164 & $166 \pm
  1$ \\ OCDM & 599 & $603 \pm 6$ & 368 & $364 \pm 3$ \\ LCDM & 593 &
  $606 \pm 10$ & 380 & $377 \pm 4$ \\ \hline
\end{tabular}
\end{center}
\end{table}
%%%%%%%%%%%%%%%%%%%%%%%%%%%%%%%%%%%%%%%%%%%%%%%%%%%%%%%%%%%%%%%%%%%%%%%%%%

As Table~\ref{tab:sigmap} indicates, the above analytical fit is in
good agreement with our simulation results.
Finally, a constraint on $\sigma_8$ and $\Omega_0$ from cluster
abundances at $z=0$ is fairly well-established. Thus combining these
model predictions and observational constraints, we will be left with
only two {\it unknown} parameters, $\Omega_0$ and $\lambda_0$, which we
desire to determine from the cosmological redshift distortion.

%%%%%%%%%%%%%%%%%%%%%%%%%%%%%%%%%%%%%%%%%%%%%%%%%%%%%%%%%%%%%%%%%%%%%
\begin{figure}[htb]
\begin{center}
   \leavevmode\epsfysize=7.1cm \epsfbox{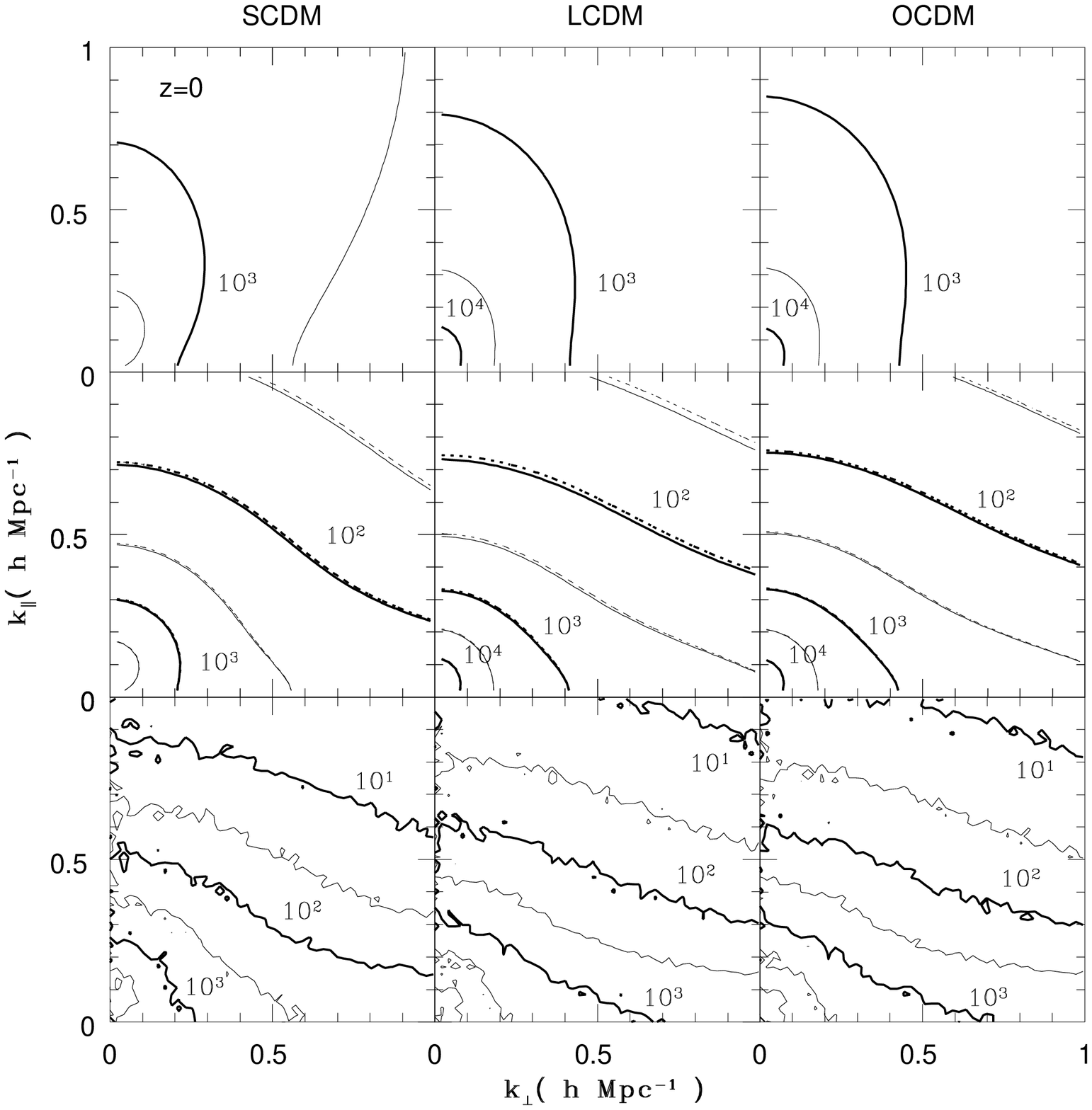}
\end{center}
\caption{ $P^{(s)}(k_{s\sperp},k_{s\spara})$ at $z=0$.
  The upper and lower panels display the linear theory predictions and
  the results from N-body simulations using all particles ($N=256^3$).
  The middle panels present our nonlinear model predictions on the basis
  of the non-linear power spectrum (Eq.~(\protect\ref{eq:pdfit}\protect)).
  For the values of the pair-wise peculiar velocity dispersion
  $\sigma_P$, the solid lines correspond to those from the simulation data, while
  the dotted lines correspond to the analytical fitting formula
  (\protect\ref{eq:spmjb}\protect).  Contour spacings are $\Delta
  \log_{10}P=0.5$, and the contours corresponding to $10^5$, $10^4$,
  $10^3$, $10^2$ and $10^1$ are plotted by the thick lines.
  \label{fig:pwcontz0}} \vspace*{0.3cm}
\begin{center}
   \leavevmode\epsfysize=7.1cm \epsfbox{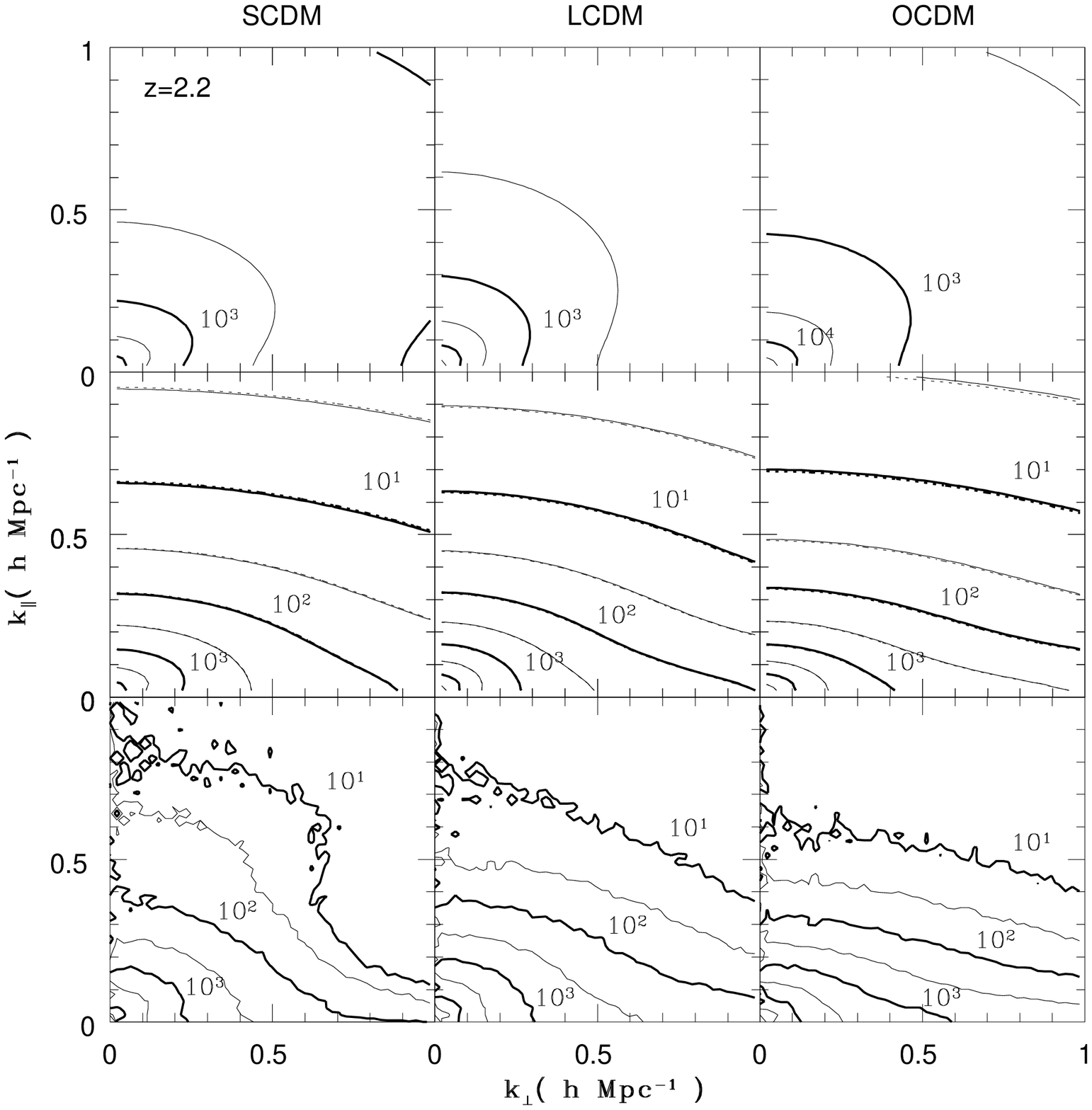}
\end{center}
\caption{The same as Fig.~\protect\ref{fig:pwcontz0}\protect,  but at
  $z=2.2$.  \label{fig:pwcontz2}}
\end{figure}
%%%%%%%%%%%%%%%%%%%%%%%%%%%%%%%%%%%%%%%%%%%%%%%%%%%%%%%%%%%%%%%%%%%%%
Figures~\ref{fig:pwcontz0} and \ref{fig:pwcontz2} display the contour plots
for $P^{(CRD)}(k_{s\sperp},k_{s\spara})$ at $z=0$ and $z=2.2$,
respectively.  The upper, middle, and lower panels correspond to
theoretical predictions in linear theory, nonlinear model predictions on
the basis of equations (\ref{eq:pdfit}) and (\ref{eq:pkcrd}), and
simulation results, respectively. Note that $P(k)$ in this section is
related to $\Delta^2(k)$ in \S \ref{subsec:predictxi} as\cite{PD94}
%%%%%%%%%%%%%%%%%%%%%%%%%%%%%%%%%%%%%%%%%%%%%%%%%%%%%%%%%%%%%%%%%%%%%%%%%%
\begin{equation}
  \Delta^2(k)=\frac{1}{(2\pi)^3}4\pi k^3P(k).
\end{equation}
%%%%%%%%%%%%%%%%%%%%%%%%%%%%%%%%%%%%%%%%%%%%%%%%%%%%%%%%%%%%%%%%%%%%%%%%%%
The middle panels plot two nonlinear models which adopt different
$\sigma_P$ in Eq. (\ref{eq:damping}). The solid curves correspond to the
pair-wise velocity dispersions directly evaluated from the simulation
data, while the dotted curves correspond to an analytical fitting formula\cite{MJB}
(Eq.~(\ref{eq:spmjb})). The right-hand-side of the above equation
depends on the scale $R$ through a spherical top-hat window function,
$W(k;R)$, while Eq. (\ref{eq:damping}) is derived on the
assumption that $\sigma_P(z)$ is scale-independent.  Note that we
adopt the velocity dispersion in comoving coordinates.  This implies
that we have to multiply the proper velocity by the conversion factor
$H_0(1+z)/H(z)$.  We adopt the value at $R=40h^{-1}$Mpc, which is the
median value of the fitting range of our analysis (see below).

%%%%%%%%%%%%%%%%%%%%%%%%%%%%%%%%%%%%%%%%%%%%%%%%%%%%%%%%%%%%%%%%%%%%%
\begin{figure}[ht]
\begin{center}
    \leavevmode\epsfysize=10cm \epsfbox{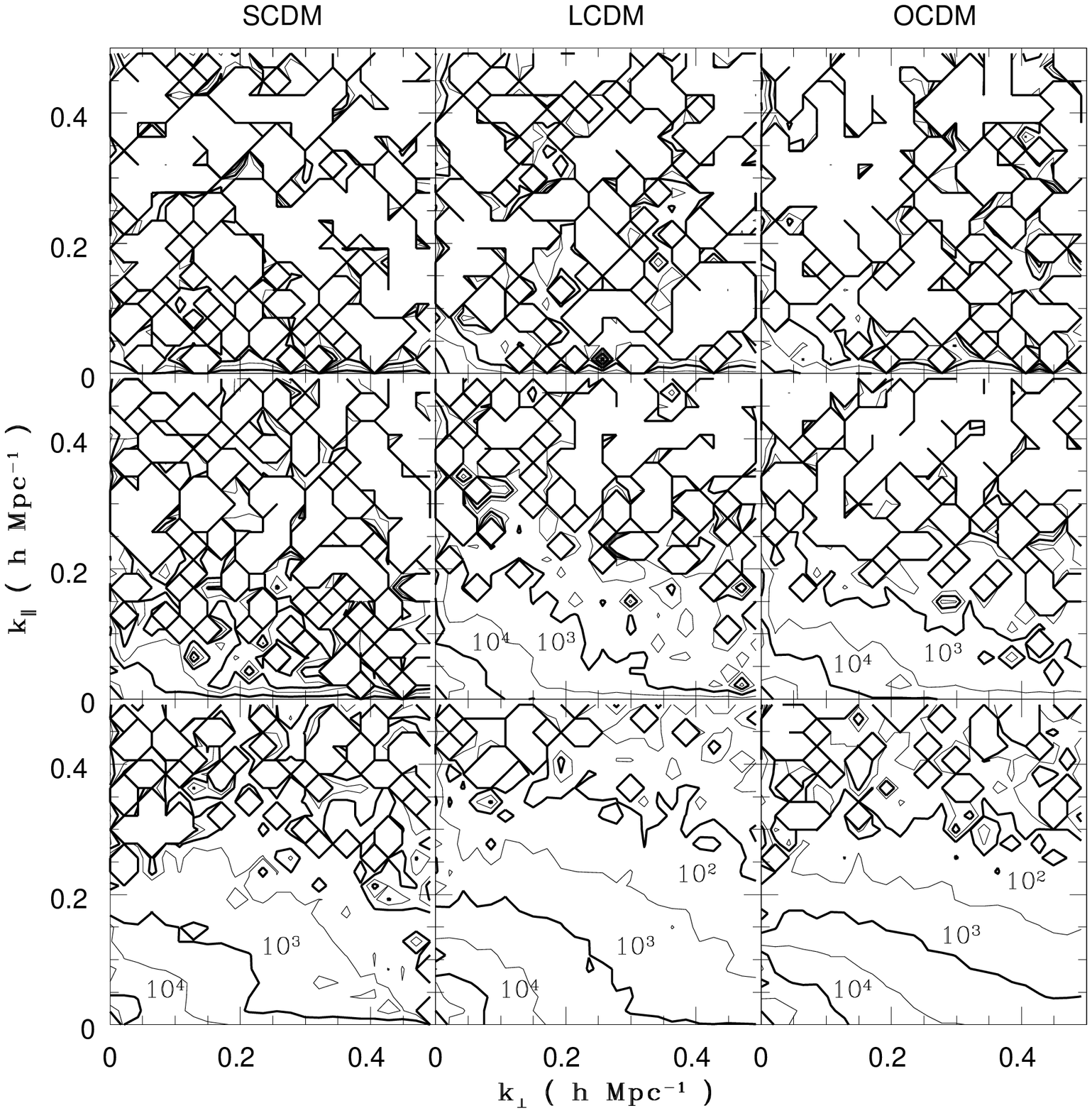}
\end{center}
\caption{ $P^{(CRD)}(k_{s\sperp},k_{s\spara})$ at $z=2.2$ from N-body
  simulations using randomly sampled particles of $5\times10^5$ ({\it
    lower}), $5\times10^4$ ({\it middle}), and $5\times10^3$ ({\it
    upper}). The contour levels are the same as those in
  Fig.~\protect\ref{fig:pwcontz0}\protect. \label{fig:pwcontcnp}}
\end{figure}
%%%%%%%%%%%%%%%%%%%%%%%%%%%%%%%%%%%%%%%%%%%%%%%%%%%%%%%%%%%%%%%%%%%%%

As in the case of the two-point correlation functions, the degree to
which one can recover the power spectrum sensitively
depends on the number of sampled particles. The lower panels in Figs.
\ref{fig:pwcontz0} and \ref{fig:pwcontz2} use all the simulation
particles.  We repeated the same analysis with randomly sampled
$N_P=5\times10^5$, $5\times10^4$ and $5\times10^3$ particles, and the
results are displayed in Fig.~\ref{fig:pwcontcnp}. The SDSS
QSO surveys, for instance, expect to have O($10^4$) QSOs between
$z=1.5$ and $2.5$. This figure implies that although the
phenomenological nonlinear models reproduce the simulation results
very well, the statistical noise due to the limited numbers of QSOs
will dominate the cosmological signal as long as one attempts to
directly compare the $P^{(CRD)}(k_{s\sperp},k_{s\spara})$.

%%%%%%%%%%%%%%%%%%%%%%%%%%%%%%%%%%%%%%%%%%%%%%%%%%%%%%%%%%%%%%%%%%%%%
\begin{figure}[tbh]
\begin{center}
    \leavevmode\epsfysize=7cm \epsfbox{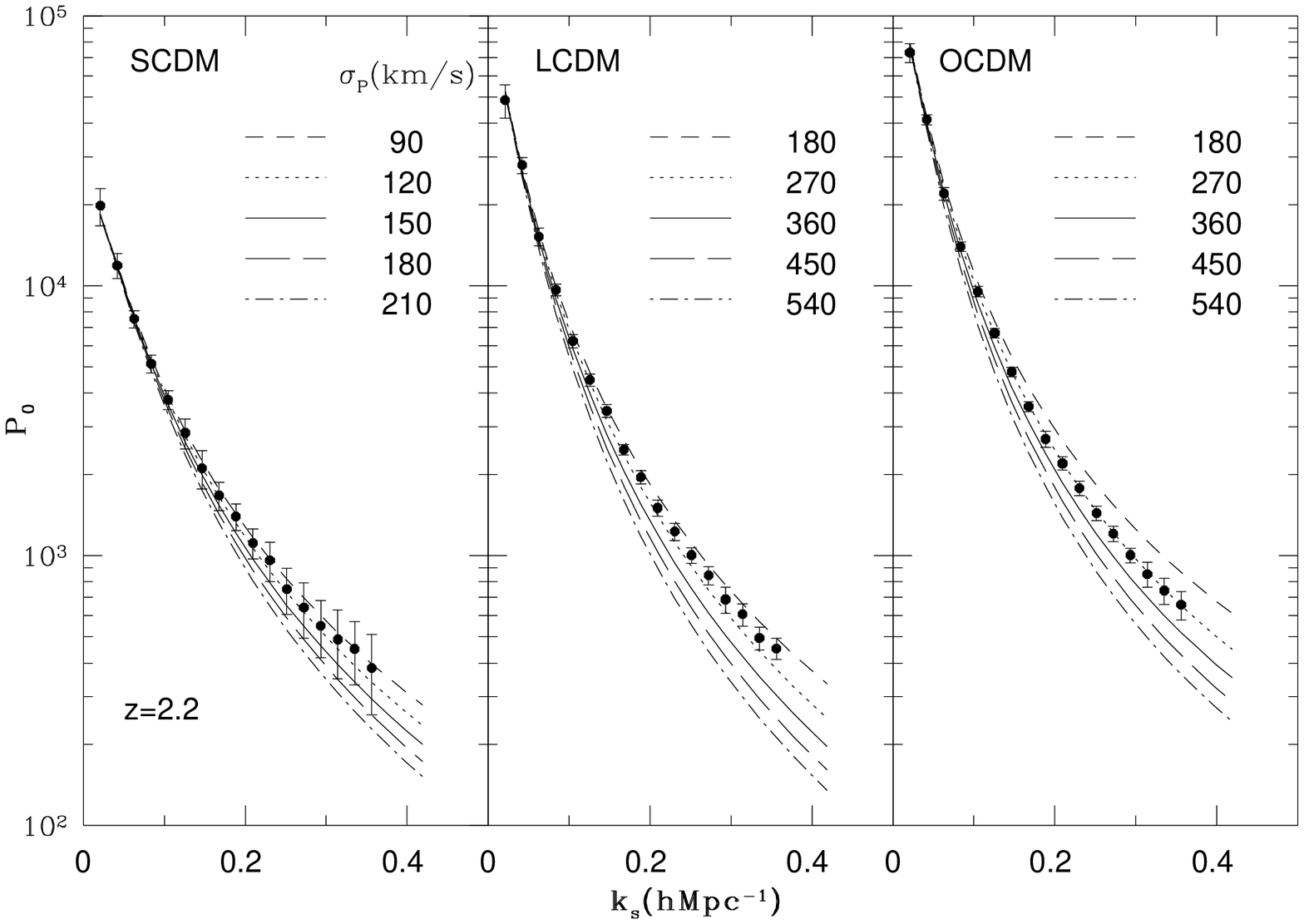}
\end{center}
\caption{ The monopole moment, $P_0^{(CRD)}(k_s)$, of the power spectrum
  in redshift space at $z=2.2$. The symbols indicate the mean values
  of 24 subsamples which randomly select $5\times10^4$ particles from
  the entire set of simulation particles in three different
  realizations. The error bars are computed from the $1\sigma$
  dispersions among the subsamples for each model. Five theoretical
  predictions are plotted with different types of curves. These values
  of $\sigma_P$ as curves correspond to various quoted in the plot.
 \label{fig:p0comb}}
\end{figure}
%%%%%%%%%%%%%%%%%%%%%%%%%%%%%%%%%%%%%%%%%%%%%%%%%%%%%%%%%%%%%%%%%%%%%

One can increase the signal-to-noise ratio by expanding the power
spectrum in multipole moments as 
%%%%%%%%%%%%%%%%%%%%%%%%%%%%%%%%%%%%%%%%%%%%%%%%%%%%%%%%%%%%%%%%%%%%%%%%%%
\begin{equation}
\label{eq:lthmoment}
  P_l^{(CRD)}(k_s;z)\equiv
\frac{2l+1}{2}\int^1_{-1}d\mu P^{(CRD)}(k_s,\mu;z)P_l(\mu),
\end{equation}
%%%%%%%%%%%%%%%%%%%%%%%%%%%%%%%%%%%%%%%%%%%%%%%%%%%%%%%%%%%%%%%%%%%%%%%%%%
where the $P_l$ are the Legendre polynomials.  In order to illustrate
the higher signal-to-noise ratio in this approach, we plot the
monopole term, $P_0^{(CRD)}(k_s;z=2.2)$, in Fig.~\ref{fig:p0comb}
computed from $5\times10^4$ randomly sampled particles.  The quoted
error bars are estimated from the $1\sigma$ dispersions of $P_0$ of 24
random subsamples in total (eight randomly sampled particle sets for
three different realizations). The five curves of different line types
correspond to theoretical predictions which use different values
for $\sigma_P$ quoted in the plot (but fix the other parameters as the
values adopted in the simulations). While the power spectrum itself is
rather noisy (see the middle panels in Fig.~\ref{fig:pwcontcnp}), the
estimated moment is very robust. Figure~\ref{fig:p0comb} suggests that
the best-fit $\sigma_P$ is systematically smaller than the values
listed in Table~\ref{tab:sigmap}. Better quantitative agreement is
obtained by replacing the $\sigma_P$ in Eq. (\ref{eq:damping}) with
the pair-wise velocity dispersion divided by $\sqrt{2}$. This is
related to the validity of the modeling of nonlinear velocity
correction (e.g., Eq.~(\ref{eq:damping})), and will be discussed in
detail elsewhere.\cite{Magira}

%%%%%%%%%%%%%%%%%%%%%%%%%%%%%%%%%%%%%%%%%%%%%%%%%%%%%%%%%%%%%%%%%%%%%
\begin{figure}[ht]
\begin{center}
    \leavevmode\epsfysize=6.0cm \epsfbox{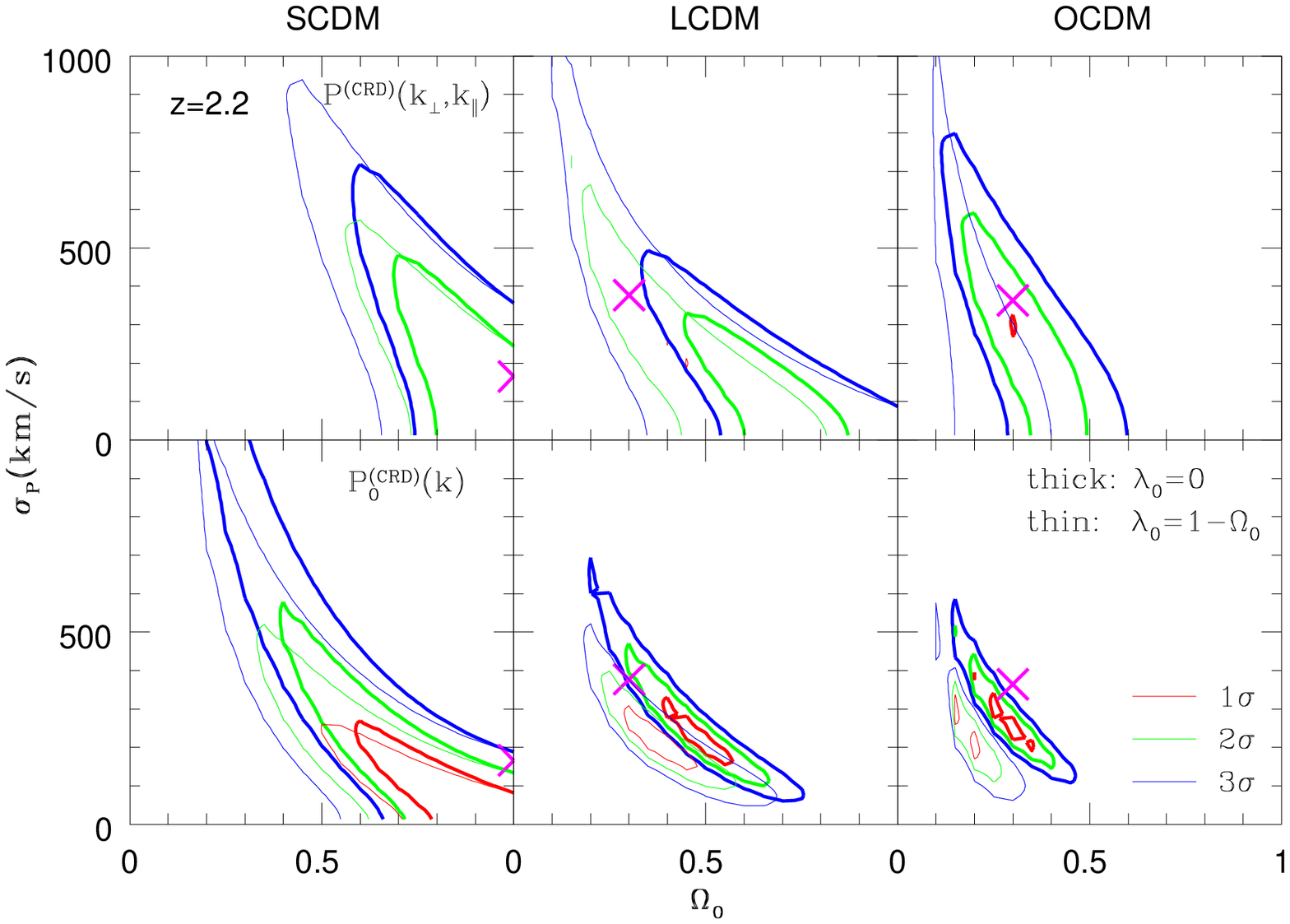}
\end{center}
\caption{ The constraints on the $\Omega_0$ - $\sigma_P$ plane derived from
 the $\chi^2$ analysis of simulation data at $z=2.2$. The upper panels are
 from $P^{(CRD)}(k_{s\sperp}, k_{s\spara})$ with all the simulation
 particles ($N=256^3$), while the lower panels are from $P_0^{(CRD)}(k_s)$
 with $5\times10^4$ randomly sampled particles.  Thick and thin contours
 correspond to the results assuming $\lambda_0=0$ and
 $\lambda_0=1-\Omega_0$, respectively.  The crosses indicate the true
 values adopted in the simulation models. We assume
 Eq.~(\protect\ref{eq:s8o0}) for the value of $\sigma_8$.
 \label{fig:chi2osp} }
\vspace*{0.5cm}
\begin{center}
    \leavevmode\epsfysize=6.0cm \epsfbox{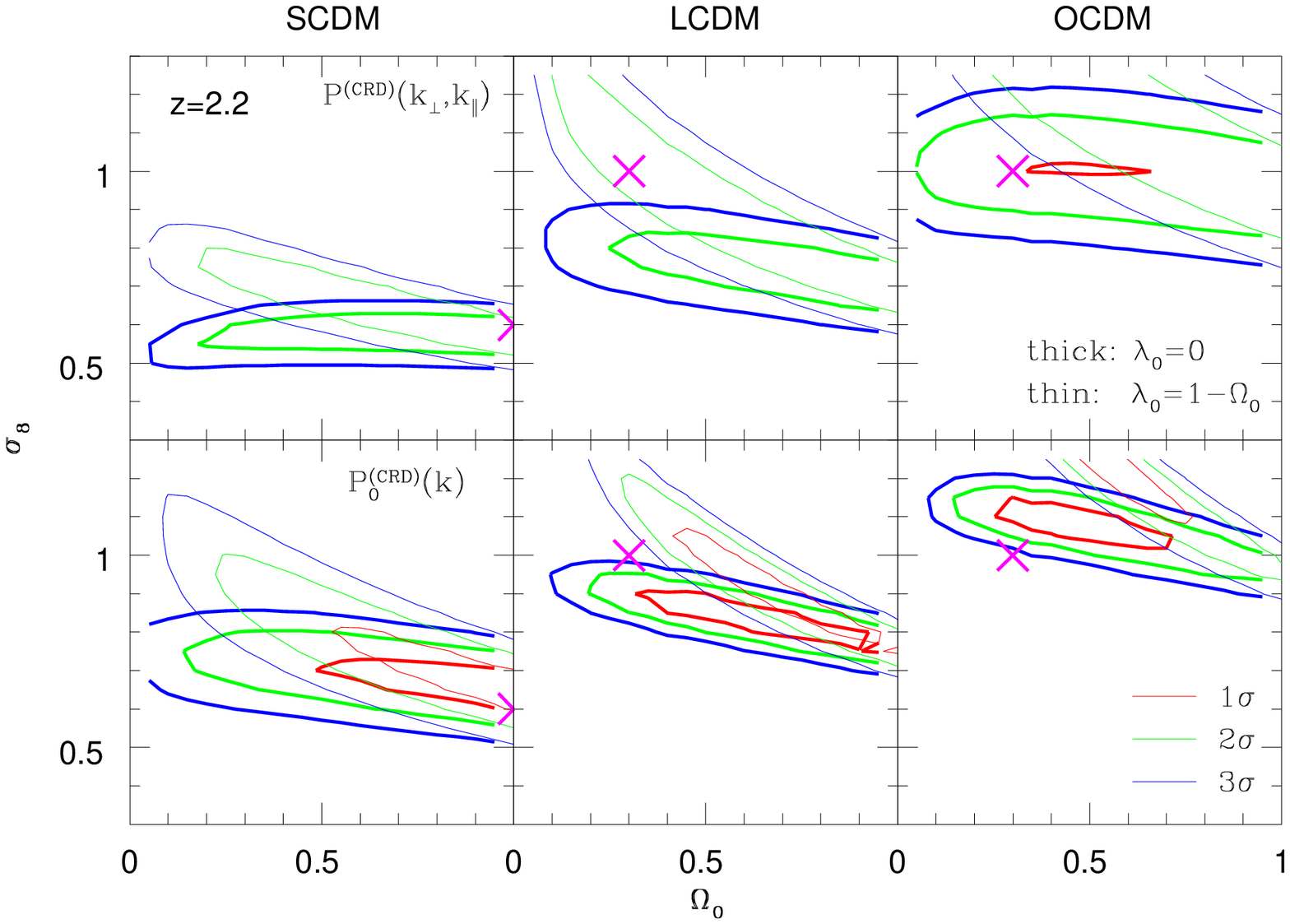}
\end{center}
\caption{ The same as Fig.~\protect\ref{fig:chi2osp}\protect, but in the
   $\Omega_0$ - $\sigma_8$ plane.  We adopt an analytical fitting
   formula (Eq.~(\protect\ref{eq:spmjb}\protect)) for the value of
   $\sigma_P$. \label{fig:chi2os8} }
\end{figure}
%%%%%%%%%%%%%%%%%%%%%%%%%%%%%%%%%%%%%%%%%%%%%%%%%%%%%%%%%%%%%%%%%%%%%

Finally we compute the $\chi^2$ between the theoretical predictions
(with nonlinear corrections) and the simulations by varying the
parameters.  The results are shown as contours in $\Omega_0$ -
$\sigma_P$ and $\Omega_0$ - $\sigma_8$ planes in
Figs.~\ref{fig:chi2osp} and \ref{fig:chi2os8}, respectively.  In these
figures, the upper and lower panels correspond to the analysis based
on $P^{(CRD)}(k_{s\sperp}, k_{s\spara})$ using all simulation
particles ($N=256^3$) and $P_0^{(CRD)}(k_s)$ using $5\times10^4$
randomly sampled particles (one realization from each model in
Table~\ref{tab:modelparam}). The $\chi^2$-fit is carried out in the
range of $(2\pi/60)h {\rm Mpc}^{-1} <k_\sperp, k_\spara, k_s<
(2\pi/20)h {\rm Mpc}^{-1}$.

In Fig.~\ref{fig:chi2osp}, we fix the value of $\sigma_8$ according to
the fitting formula based on the following cluster abundances:\cite{KS}
%%%%%%%%%%%%%%%%%%%%%%%%%%%%%%%%%%%%%%%%%%%%%%%%%%%%%%%%%%%%%%%%%%%
\begin{eqnarray}
  \sigma_8 = (0.54 \pm 0.02 ) \times \left\{
      \begin{array}{ll}
        \Omega_0^{-0.35-0.82\Omega_0+0.55\Omega_0^2} &
        \mbox{($\lambda_0=1-\Omega_0$)}, \\ 
        \Omega_0^{-0.28-0.91\Omega_0+0.68\Omega_0^2} &
        \mbox{($\lambda_0=0$)} . 
      \end{array}
   \right. 
\label{eq:s8o0}
\end{eqnarray}
%%%%%%%%%%%%%%%%%%%%%%%%%%%%%%%%%%%%%%%%%%%%%%%%%%%%%%%%%%%%%%%%%%%
In Fig.~\ref{fig:chi2os8}, we fix the value of $\sigma_P$ according to
Eq. (14) in Ref.~\citen{MJB}. Incidentally the cluster abundance
constraints (\ref{eq:s8o0})  are fairly orthogonal to our
constraints from the redshift distortion.  

The best-fit values for $\sigma_P$ and $\sigma_8$ in the above plots
are slightly smaller than their true values (marked as crosses). This is
related to the nonlinear velocity correction, as described above, and
we can easily correct for this systematic effect by adopting a more
appropriate model.\cite{Magira} Therefore we conclude that it is
feasible to break the degeneracy in the cosmological parameters by
combining the cosmological redshift-space distortion in the future QSO
samples with other cosmological tests, despite the fact that the
present modeling of nonlinear effects is fairly empirical.

\section{Conclusion \label{sec:final}}

The present paper focuses on two important effects, the cosmological
light-cone effect and redshift-space distortion, which have been largely
ignored in previous discussions of clustering statistics.  We have
demonstrated that they play an important role in the analysis of the
upcoming redshift surveys, particularly of high-redshift objects, as
both cosmological signals and noise depending on specific aspects of
the phenomena that one is interested in. We summarize our main
conclusions here.
%%%%%%%%%%%%%%%%%%%%%%%%%%%%%%%%%%%%%%%%%%%%%%%%%%%%%%%%%%%%%%%%%%%%%%
\begin{enumerate}
\item We derived an expression for the two-point correlation function
  properly defined on the light-cone hypersurface.\cite{YS} This
  expression is easily evaluated numerically when the underlying
  cosmological model is specified.  With this, one can directly
  confront the resulting predictions with the observational data in a
  fairly straightforward manner.
\item The cosmological light-cone effect produces artificial
  scale-dependence and redshift-dependence on the higher-order
  moments\cite{MSS} of redshift-space clustering of any cosmological
  objects.
\item In linear theory we formulated the cosmological redshift-space
  distortion\cite{MS96} which induces an apparent anisotropy in
  two-point correlation functions, especially at high redshifts.
  Further detailed studies with N-body simulations\cite{Magira}
  indicate that it is feasible to constrain the cosmological
  parameters from the future QSO samples via this effect even though
  the nonlinear evolution appreciably affects the linear theory
  predictions.
\end{enumerate}
%%%%%%%%%%%%%%%%%%%%%%%%%%%%%%%%%%%%%%%%%%%%%%%%%%%%%%%%%%%%%%%%%%%%%%

Apparently, the results described above should be regarded as
the first attempts to raise the importance and basic features of these
two cosmological effects. These results are still far from complete in the sense
that there are many aspects which remain to be explored. We hope that
this short review serves as a practical and useful introductory note
for more detailed investigations in the future.

\section*{Acknowledgments}

We thank Kenji Tomita for inviting us to write this article and for
useful comments on the manuscript.  The materials presented in \S
\ref{subsec:highlc} and \S \ref{subsec:betacrd} are based on our
collaborative work with Istv\'an Szapudi and Takahiro T.Nakamura,
respectively, whom we thank for useful discussions.  We thank Saavik
K. Ford for a careful reading of the manuscript.  T.M. and Y.P.J
gratefully acknowledge the fellowship from the Japan Society for the
Promotion of Science.  Numerical computations presented in \S
\ref{subsec:nbodycrd} were carried out on VPP300/16R and VX/4R at ADAC
(the Astronomical Data Analysis Center) of the National Astronomical
Observatory, Japan, as well as at RESCEU (Research Center for the
Early Universe, University of Tokyo) and KEK (National Laboratory for
High Energy Physics, Japan). This research was supported in part by
the Grants-in-Aid by the Ministry of Education, Science, Sports and
Culture of Japan (07CE2002) to RESCEU, and by the Supercomputer
Project (No.98-35) of High Energy Accelerator Research Organization
(KEK).


\begin{thebibliography}{99}
%%%%%%%%%%%%%%%%%%%%%%%%%%%%%%%%%%%%%%%%%%%%%%%%%%%%%%%%%%%%%
% Some macros are available for the bibliography:
%   o for general use
%      \JL : general journals          \andvol : Vol (Year) Page
%   o for individual journal 
%      \PR  : Phys. Rev.               \PRL : Phys. Rev. Lett.
%      \NP  : Nucl. Phys.              \PL  : Phys. Lett.
%      \JMP : J. Math. Phys.           \CMP : Commun. Math. Phys.
%      \PTP : Prog. Theor. Phys.       \JPSJ: J. Phys. Soc. Jpn.
%      \JP  : J. of Phys.              \NC  : Nouvo Cim.
%      \IJMP: Int. J. Mod. Phys.       \ANN : Ann. of Phys.
% Usage:
%   \PR{D45,1990,345}            ==> Phys.~Rev.\ {\bf D45} (1990), 345
%   \JL{Phys.~Lett.,A30,1981,56} ==> Phys.~Lett.\ {\bf A30} (1981), 56
%   \andvol{B123,1995,1020}      ==> {\bf B123} (1995), 1020
%%%%%%%%%%%%%%%%%%%%%%%%%%%%%%%%%%%%%%%%%%%%%%%%%%%%%%%%%%%%%
\bibitem{GH} M.~J.~Geller and J.~P.~Huchra,  Science \ {\bf 246} (1989), 897.
\bibitem{Steidel} C.~C.~Steidel, K.~L.~Adelberger, M.~Dickinson,
M.~Giavalisco, M.~Pettini and M.~Kellogg, \JL{Astrophys.J.,492,1998,428}.
\bibitem{Carrera} F.~J.~Carrera et~al., \JL{MNRAS,299,1998,229}.
\bibitem{Boyle} B.~J.~Boyle, S.~M.~Croom, R.~J.~Smith, T.~Shanks,
L.~Miller and N.~Loaring, Phil.Trans.R.Soc.Lond.A \ (1998), in press
(astro-ph/9805140).
\bibitem{YS} K.~Yamamoto and Y.~Suto,\JL{Astrophys.J.,517,1999,in press 
(astro-ph/9812486)}.
\bibitem{MSS} T.~Matsubara, Y.~Suto and I.~Szapudi, 
\JL{Astrophys.J.,491,1997,L1}.
\bibitem{MS96} T.~Matsubara and Y.~Suto, \JL{Astrophys.J.,470,1996,L1}.
\bibitem{NMS} T.~T.~Nakamura, T.~Matsubara and Y.~Suto, 
\JL{Astrophys.J.,414,1998,13}.
\bibitem{KS} T.~Kitayama and Y.~Suto, \JL{Astrophys.J.,490,1997,557}.
\bibitem{BBKS} J.~M.~Bardeen, J.~R.~Bond, N.~Kaiser and A.~S.~Szalay, 
\JL{Astrophys.J.,304,1985,15}.
\bibitem{PD94} J.~A.~Peacock and S.~J.~Dodds, \JL{MNRAS,267,1994,1020}.
\bibitem{PD96} J.~A.~Peacock and S.~J.~Dodds, \JL{MNRAS,280,1996,L19}.
\bibitem{Fry96} J.~N.~Fry, \JL{Astrophys.J.,461,1996, L65}.
\bibitem{WN} S.~Wallington and R.~Narayan, \JL{Astrophys.J.,403,1993,517}.
\bibitem{NS} T.~T.~Nakamura and Y.~Suto, \PTP{97,1997,49}.
\bibitem{FAC} F.~La Franca, P.~Andreani and S.~Cristiani, 
\JL{Astrophys.J.,497,1998,529}.
\bibitem{Mataresse} S.~Matarrese, P.~Coles, F.~Lucchin and L.~Moscardini,
 \JL{MNRAS,286,1997,115}.
\bibitem{Fry84} J.~N.~Fry, \JL{Astrophys.J.,277,1984,L5}.
\bibitem{Bouchet} F.~R.~Bouchet, R.~Juszkiewicz, S.~Colombi and R.~Pellat, 
\JL{Astrophys.J.,394,1992,L5}.
\bibitem{JMW} B.~Jain, H.~J.~Mo and S.~D.~M.~White, \JL{MNRAS,276,1995,L25}.
\bibitem{AP} C.~Alcock and B.~Paczy\'nski, \JL{Nature,281,1979,358}.
\bibitem{BPH} W.~E.~Ballinger, J.~A.~Peacock and A.~F.~Heavens, 
\JL{MNRAS,282,1996,877}.
\bibitem{DP83} M.~Davis and P.~J.~E.~Peebles, \JL{Astrophys.J.,267,1983,465}.
\bibitem{Kaiser87} N.~Kaiser, \JL{MNRAS,227,1987,1}.
\bibitem{Hamilton92} A.~J.~S.~Hamilton, \JL{Astrophys.J.,385,1992,L5}.
\bibitem{Hamilton97} A.~J.~S.~Hamilton, to appear in the Proceedings of
  Ringberg Workshop on Large-Scale Structure, ed. D.~Hamilton \ 
  (1997), (astro-ph/9708102).
\bibitem{Magira} H.~Magira, Y.~P.~Jing and Y.~Suto (1999), in preparation
\bibitem{Peebles93} P.~J.~E.~Peebles,  Principles of Physical Cosmology
  (Princeton: Princeton Univ. Press, 1993).
\bibitem{GW} J.~E.~Gunn and D.~H.~Weinberg, in Wide Field
  Spectroscopy and the Distant Universe, ed. S.~J.~Maddox and
  A.~Ara\'gon-Salamanca (World Scientific, Singapore, 1995).
\bibitem{DR73} C.~C.~Dyer and R.~C.~Roeder, \JL{Astrophys.J.,180,1973,L31}.
\bibitem{tomita} K.~Tomita, \PTP{100,1998,1}.
\bibitem{JS98} Y.~P.~Jing and Y.~Suto, \JL{Astrophys.J.,494,1998,L5}.
\bibitem{Jing} Y.~P.~Jing, \JL{Astrophys.J.,503,1998,L9}.
\bibitem{MJB} H.~J.~Mo, Y.~P.~Jing and G.~B\"{o}rner \JL{MNRAS,286,1997,979}.
\bibitem{BE} C.~M.~Baugh and G.~Efstathiou \JL{MNRAS,267,1994,323}.

\end{thebibliography}
\end{document}